\documentclass[12pt]{article}

\usepackage{amsmath} 
\allowdisplaybreaks
\usepackage{amsfonts} 
\usepackage{mathtools} 
\usepackage{amssymb} 
\usepackage{mathrsfs} 
\usepackage[mathscr]{euscript} 
\usepackage[a4paper,margin=1.0in]{geometry} 
\usepackage{graphicx,graphics,subfig} 
\usepackage{microtype}
\usepackage{float}
\usepackage[colorlinks=true,citecolor=DarkGreen,linkcolor=DarkRed,urlcolor=DarkBlue,menucolor=Blue,anchorcolor=DarkRed,
filecolor=Black,pdfauthor=author,linktocpage,pdfpagelabels,pdfborder={0 0 0},hyperfootnotes=false,bookmarks=true,
bookmarksnumbered=true]{hyperref} 
\usepackage{cleveref}
\usepackage{algorithm}
\usepackage{setspace} 
\usepackage{latexsym} 
\usepackage[T1]{fontenc} 
\usepackage[square,comma,numbers,sort&compress]{natbib}
\usepackage[utf8]{inputenc} 
\usepackage[english]{babel} 
\usepackage[margin=10pt,font={small,stretch=1.0},labelfont=bf]{caption} 
\usepackage[usenames,dvipsnames,svgnames,x11names,table]{xcolor} 
\usepackage{tabularx} 
\usepackage{enumerate} 
\usepackage{expdlist} 
\usepackage{psfrag} 
\usepackage{physics}
\usepackage{epsfig}
\usepackage{epstopdf}

\usepackage{soul} 
\sethlcolor{yellow}
\setstcolor{red}
\setulcolor{green}
\usepackage[normalem]{ulem} 

\usepackage[titles,subfigure]{tocloft}
\hypersetup{linktoc=all}

\makeatletter
\renewcommand{\cftsecpagefont}{\HyColor@HyperrefColor{DarkRed}\@linkcolor}
\renewcommand{\cftsubsecpagefont}{\HyColor@HyperrefColor{DarkRed}\@linkcolor}
\renewcommand{\cftsubsubsecpagefont}{\HyColor@HyperrefColor{DarkRed}\@linkcolor}
\makeatother

\makeatletter \@addtoreset{equation}{section} \makeatother

\begin{document}

\begin{titlepage}
\hypersetup{pageanchor=false}

\renewcommand{\thefootnote}{\textcolor{DarkGreen}{\fnsymbol{footnote}}}

\begin{center}

\vspace*{1cm}

{\LARGE \textbf{Spinor Walls in Five-Dimensional \\ \, \\ Warped Spacetime}}

\vskip 20mm

{\large
Zheng-Quan Cui$^{a,b,c,}$\footnote{Email: \texttt{cuizhq5@mail.sysu.edu.cn}} and
Yu-Xiao Liu$^{a,b,}$\footnote{Email: \texttt{liuyx@lzu.edu.cn}, corresponding author}
}

\vskip 5mm
{
$^{a}$ \textit{Institute of Theoretical Physics $\&$ Research Center of Gravitation, Lanzhou University,
Lanzhou 730000, China}\\
$^{b}$ \textit{Lanzhou Center for Theoretical Physics, Key Laboratory of Theoretical Physics of Gansu Province, Lanzhou University, Lanzhou 730000, China}\\
$^{c}$ \textit{Centre for Particle Theory, Department of Mathematical Sciences, Durham University,
Durham DH1 3LE, UK}
}

\vskip 5mm

\today

\end{center}

\vskip 10mm

\begin{center}
{\textbf{\textsc{Abstract}}}
\end{center}
We study domain wall solutions of a real spinor field coupling with gravitation in five dimensions. We find that the nonlinear spinor field supports a class of soliton configurations which could be viewed as a wall embedded in five dimensions. We begin with an illuminating solution of the spinor field in the absence of gravitation. In a further investigation, we exhibit three sets of solutions of the spinor field with nonconstant curvature bulk spacetimes and three sets of solutions corresponding to three constant curvature bulk spacetimes. We demonstrate that some of these solutions in specific conditions have the energy density distributions of domain walls for the spinor field, where the scalar curvature is regular everywhere. Therefore, the configurations of these walls can be interpreted as spinor walls which are interesting spinor field realizations of domain walls. In order to investigate the stability of these spinor configurations, the linear perturbations are considered. The localization of the zero mode of tensor perturbation is also discussed.

\vskip 15mm


\end{titlepage}

\newpage
\hypersetup{pageanchor=true}
\pagenumbering{arabic} 

\renewcommand*{\thefootnote}{\arabic{footnote}}
\setcounter{footnote}{0}

\begin{spacing}{1.2}
\hypersetup{linkcolor=Black,filecolor=DarkGreen,urlcolor=DarkBlue}
\tableofcontents
\end{spacing}

\section{Introduction}

Many nonlinear physical systems support soliton solutions, which represent spatially localized field configurations. Kinks and domain walls are a subclass of co-dimension $1$ topological solitons. Kinks are stable configurations interpolating between two of more degenerate vacua in $(1 + 1)$-dimensional spacetime, while domain walls are the higher-dimensional analogues of kinks. They have the following properties: (i) the corresponding solutions are static and depend only on one spatial coordinate; (ii) they are topologically stable and indestructible --- once a kink or domain wall is created it cannot spontaneously disappear. They are invaluable for nonperturbative aspects of field theories. A five-dimensional domain wall model in the absence of gravitation was presented in Ref.~\cite{Rubakov:1983rsd}, where fermions can be confined on the wall by the Yukawa coupling between fermions and the background scalar field generating the domain wall. Kinks and domain walls in the presence of gravitation are more remarkable for extra dimension scenarios, especially for brane world frameworks. In the celebrated Randall-Sundrum (RS) II model, gravitation can be localized in the vicinity of a brane, a subspace embedded in five-dimensional anti-de Sitter (AdS) spacetime~\cite{Randall:1999rsa}. It was shown that the zero mode of the graviton is localized on the delta-function-like brane and is responsible for the Newtonian potential, while the massive Kaluza-Klein (KK) modes of the graviton make corrections to the Newtonian potential. However, in the RS II model, the bulk scalar curvature is inevitably divergent at the location of the brane, and the extrinsic curvature satisfies the Israel junction condition. Thick branes with smooth warped extra dimensions have no such curvature singularity and could be viewed as domain walls in the presence of gravitation. For previous works on thick branes see Refs.~\cite{Gremm:2000gf,Gremm:2000gt,
DeWolfe:2000dfgk,Csaki:2000cehs,Kobayashi:2002kks,Giovannini:2001gg,Giovannini:2002ga,Giovannini:2002gb,Giovannini:2003g} and recent reviews on thick branes see~\cite{Dzhunushaliev:2010dfm,Liu:2017l,Ahluwalia:2022ttu}.

There are many thick brane models constructed by one or more scalar fields, which have achieved some features, such as the localization of matter fields, and the Newtonian potential recovered on the brane. The key point of using scalar fields is that a soliton configuration could be generated. Similarly, there are soliton solutions of spinor fields~\cite{Lee:1975lkg,Merwe:1977m,Takahashi:1979t,Nogami:1992nt,Cooper:2010ckms}. A natural question is whether spinor fields could be used to construct thick branes. It is known that the equations of motion of ordinary spinor fields without self-interactions are linear and can not produce a soliton configuration. Therefore, nonlinear spinor fields dominate the study of such kinds of solitons. Spinor fields including nonlinear ones are widely used in many fields mainly thanks to their specific behaviour in the presence of the gravitational field. Specifically, spinor fields were explored for particle-like solutions~\cite{Finster:1999fsy}, and nonlinear spinor fields were explored to explain dark energy in the quintom scenario~\cite{Cai:2010cssx,Li:2011llww}. In addition, as one of the most promising theories for unifying gravitation and other fundamental interactions, string theory at a low energy limit includes the contribution of torsion~\cite{Shapiro:2002s}. Nonlinear spinor fields might represent the existence of spacetime torsion (a geometrical entity as spacetime curvature), which is suggested by Einstein-Cartan theory~\cite{Hehl:1976hhk}. An accepted fact is that torsion does not exist in our current universe. Even though it does, it will be extremely small. To interpret a torsion-free brane, the authors of Ref.~\cite{Mukhopadhyaya:2002mss} proposed an approach to create the illusion of a torsion-free universe from a bulk with torsion. Therefore, nonlinear spinor fields in extra dimension scenarios might provide some ways to shed light on where the torsion of the world is.

Nonlinear spinor fields were early investigated in Refs.~\cite{Finkelstein:1951flr,Finkelstein:1956ffk}, for many further explorations see the review~\cite{Fushchich:1989fz}. Heisenberg studied the possibility of nonlinear spinor fields as a theory of elementary particles~\cite{Heisenberg:1953h,Heisenberg:1957h}. G\"{u}rsey proposed a spinor field equation which is similar to Heisenberg's nonlinear generalization of the Dirac equation~\cite{Gursey:1956g}. This equation is the first nonlinear conformal invariant wave equation which is viewed ever as a possible basis for a unitary description of elementary particles. An example of a self-interacting spinor field, governed by a nonlinear field equation, is the G\"{u}rsey equation~\cite{Gursey:1956g}
\begin{equation*}
   \gamma^{\mu} \partial_{\mu} \psi+\sigma(\bar{\psi} \psi)^{\frac{1}{3}} \psi=0\,,
\end{equation*}
which can be derived from the Lagrangian density
\begin{equation*}
  \mathcal{L}= -\frac{1}{2}\left(\bar{\psi} \gamma^{\mu} \partial_{\mu} \psi-\partial_{\mu}\bar{\psi}\gamma^{\mu} \psi\right)-\frac{3}{4} \sigma\left(\bar{\psi} \psi\right)^{\frac{4}{3}}\,.
\end{equation*}
Here, $\sigma$ is the coupling strength to measure the G\"{u}rsey self-interaction. In order to reserve the conformal invariance of the Lagrangian density, a nonpolynomial form of nonlinear spinor fields was used to write this Lagrangian density. For our interest, the conformal invariance of this Lagrangian density is not required. Nevertheless, we allow the existence of nonlinear spinor fields. The solutions to G{\"u}rsey's equation were studied~\cite{Kortel:1956k}. In this paper, we will investigate soliton solutions in the G\"{u}rsey-like equation.

Elementary fermions mathematically described by spinors are an essential component of our world. In previous studies on brane worlds, spinor fields were introduced for constructing thick branes in Refs.~\cite{Dzhunushaliev:2011df,Dzhunushaliev:2012df}, but there is no anticipative warp of the bulk, namely gravitation can not be localized on the brane. We attempt to explore the implementation of gravitational domain walls of spinor fields by extracting insights from domain walls generated by scalar fields. In our exploration, nonlinear spinor fields are considered and the equations of motion are G\"{u}rsey-like equations.

This paper is organized as follows. In section~\ref{sec:review}, we give a brief review of spinor fields in five-dimensional spacetime. In section~\ref{sec:solutions_flat}, we introduce a toy model for a spinor field in five-dimensional flat spacetime. In section~\ref{sec:model}, analytic solutions of spinor fields are studied both in nonconstant curvature and constant curvature spacetimes. In section~\ref{sec:stability}, we investigate the stability of the spinor walls. In section~\ref{sec:localization}, the zero modes of tensor perturbations are discussed. Finally, discussion and conclusions are given in section~\ref{sec:conclusions}.

Throughout the paper, capital Latin letters $M,N,\ldots$ represent the five-dimensional coordinate indices running over $0, 1, 2, 3, 5$, and lower-case Greek letters $\mu,\nu,\ldots$ represent the four-dimensional coordinate indices running over $0, 1, 2, 3$. We are using the definitions $R^P_{\;MQN}=\partial_Q\Gamma^P_{\;MN}
-\partial_N\Gamma^P_{\;MQ}+\Gamma^P_{\;QL}\Gamma^L_{\;MN}-\Gamma^P_{\;NL}\Gamma^L_{\;MQ}$ and $R_{MN}=R^L_{\;MLN}$. The metric signature is $(-,+,+,+,+)$. We use the units with $8\pi G_{\text{N}}^{(5)}=c=\hbar=1$, where $G_{\text{N}}^{(5)}$ is the five-dimensional gravitational constant.

\section{Spinor fields in five-dimensional spacetime}
\label{sec:review}

We firstly review the theory of spinor fields in five-dimensional spacetime. There are many different conventions in use for spinor fields. Our conventions about spinor fields are closest to those in Ref.~\cite{Weinberg:1995w}. Introduction to spinor fields refers to Refs.~\cite{Birrell:1982bd,Parker:2009ptq}.

In general, gamma matrices are introduced to construct generators of Lorentz group in spinor representation and can be chosen in any representation. Considering the chiral nature of spinor fields, the gamma matrices which we adopt in five-dimensional flat spacetime are
\begin{align}\label{eq:5gamma}
  &\gamma^{\bar{0}}=-\mathrm{i}
\begin{pmatrix}
0 & I  \\
I & 0
\end{pmatrix} \,, \quad
  \gamma^{\bar{i}}=-\mathrm{i}
\begin{pmatrix}
0 & \sigma^{\bar{i}}  \\
-\sigma^{\bar{i}} & 0
\end{pmatrix} \,, \quad
  \gamma^{\bar{5}}=
\begin{pmatrix}
I & 0  \\
0 & -I
\end{pmatrix} \,,
\end{align}
where the indices with $\;\bar{}\;$ are the ones of flat spacetime coordinates, $I$ is a $2\times2$ unity matrix, and $\sigma^{\bar{i}}$ are Pauli matrices with $\bar{i}=1, 2, 3$. These gamma matrices satisfy the Clifford algebra $\{\gamma^{\bar{M}},\gamma^{\bar{N}}\}=2\eta^{\bar{M}\bar{N}}$ with $\eta_{\bar{M}\bar{N}}=\mathrm{diag}(-1,1,1,1,1)$. The action of a spinor field in flat spacetime reads
\begin{equation}\label{eq:spinor_action_flat}
S_\psi=\int\mathrm{d}^5x\left[-\frac{1}{2}\left(\bar{\psi}\gamma^{\bar{M}}\partial_{\bar{M}}\psi
-\partial_{\bar{M}}\bar{\psi}\gamma^{\bar{M}}\psi\right)-V(\bar{\psi},\psi)\right]\,,
\end{equation}
where $\bar{\psi}=\mathrm{i}\psi^\dagger\gamma^{\bar{0}}$ is the conjugation of the spinor $\psi$ which generally has four complex components in five-dimensional spacetime. This potential term may include self-interactions of the spinor field. The equations of motion for the spinor field are
\begin{align}\label{eq:spinor_EOM_flat}
  \gamma^{\bar{M}}\partial_{\bar{M}}\psi+\frac{\partial V}{\partial \bar{\psi}}=0\,, \quad \partial_{\bar{M}}\bar{\psi}\gamma^{\bar{M}}-\frac{\partial V}{\partial \psi}=0\,.
\end{align}
The energy-momentum tensor (on-shell), i.e., the Belinfante tensor, for the spinor field in the action~\eqref{eq:spinor_action_flat} is arrived at~\cite{Ortin:2015o}
\begin{align}\label{eq:EM_tensor_flat}
    T_{\bar{N}}^{\bar{M}}= &\frac{1}{4}\left[\bar{\psi}\left(\gamma^{\bar{M}}\eta^{\bar{L}}_{\bar{N}}
    +\gamma_{\bar{N}}\eta^{\bar{M}\bar{L}}\right)\partial_{\bar{L}}\psi
  -\partial_{\bar{L}}\bar{\psi}\left(\gamma^{\bar{M}}\eta^{\bar{L}}_{\bar{N}}
    +\gamma_{\bar{N}}\eta^{\bar{M}\bar{L}}\right)\psi\right] \nonumber \\  &+\eta_{\bar{N}}^{\bar{M}}\left[\frac{1}{2}\left(\bar{\psi}\frac{\partial V}{\partial \bar{\psi}}
   +\frac{\partial V}{\partial \psi}\psi\right)-V\right] \,.
\end{align}

It is well known that there are no spinor representations of the general linear group. A local tangent space should introduce local inertial frames, vielbeins, which are necessary to introduce spinor fields in curved spacetime. The relation between the spacetime metric and the tangent space metric is
\begin{equation}\label{eq:vielbeins}
  g_{MN}=e_{\;M}^{\bar{M}} e_{\;N}^{\bar{N}}\eta_{\bar{M}\bar{N}}, \quad \eta_{\bar{M}\bar{N}}=e^{\;M}_{\bar{M}} e^{\;N}_{\bar{N}}g_{MN}\,,
\end{equation}
where the indices with and without $\;\bar{}\;$ are the local Lorentz frame indices and the general coordinate indices, respectively; $e_{\;M}^{\bar{M}}=\frac{\partial x^{\bar{M}}}{\partial x^M}$ are the vielbeins. They also satisfy the orthogonality conditions
\begin{align}\label{eq:orthogonal}
  e_{\;M}^{\bar{M}}e^{\;M}_{\bar{N}}=\delta^{\bar{M}}_{\bar{N}}\,, \quad e_{\;M}^{\bar{M}}e^{\;N}_{\bar{M}}=\delta^{N}_{M}\,.
\end{align}
The gamma matrices in curved spacetime are related to the ones in flat spacetime through the vielbeins as
\begin{align}\label{eq:gamma_matrices}
  \gamma^M=e^{\;M}_{\bar{M}}\gamma^{\bar{M}}, \quad \gamma_M=e_{\;M}^{\bar{M}}\gamma_{\bar{M}}\,.
\end{align}
They satisfy the Clifford algebra $\{\gamma^M,\gamma^N\}=2g^{MN}$. The covariant derivative of a spinor field is given by
\begin{equation}\label{eq:co_derivative}
  \nabla_M\psi=\left(\partial_M+\frac{\mathrm{i}}{2}\omega_M^{\bar{A}\bar{B}}S_{\bar{A}\bar{B}}\right)\psi\,,\quad
  \nabla_M\bar{\psi}=\bar{\psi}\left(\overleftarrow{\partial}_M-
  \frac{\mathrm{i}}{2}\omega_{M}^{\bar{A}\bar{B}}S_{\bar{A}\bar{B}}\right)\,.
\end{equation}
Here $\bar{\psi}=\mathrm{i}\psi^\dagger\gamma^0$ is the conjugation of the spinor $\psi$ with four complex components in five-dimensional curved spacetime; $\omega^{\bar{A}\bar{B}}_{M}$ can be explicitly written in terms of the vielbeins as~\cite{Green:1987gsw}
\begin{align}\label{eq:spin_connection}
  \omega^{\bar{A}\bar{B}}_{M}=&
  \frac{1}{2}e^{N\bar{A}}\left(\partial_{M}e^{\bar{B}}_{N}-\partial_{N}e^{\bar{B}}_{M}\right)
 -\frac{1}{2}e^{N\bar{B}}\left(\partial_{M}e^{\bar{A}}_{N}-\partial_{N}e^{\bar{A}}_{M}\right) \nonumber \\
 &-\frac{1}{2}e^{P\bar{A}}e^{Q\bar{B}}\left(\partial_{P}e_{Q\bar{C}}-\partial_{Q}e_{P\bar{C}}\right)e^{\bar{C}}_{M}\,,
\end{align}
and $S_{\bar{A}\bar{B}}$, the generator of spinor representation of Lorentz group, is
\begin{equation}\label{eq:generator}
  S_{\bar{A}\bar{B}}=-\frac{\mathrm{i}}{4}\left[\gamma_{\bar{A}},\gamma_{\bar{B}}\right]
  =\frac{1}{2}\Sigma_{\bar{A}\bar{B}}\,,
\end{equation}
where $\Sigma_{\bar{A}\bar{B}}$ is the Pauli operator.

The action for a spinor field minimally coupling with gravitation reads
\begin{equation}\label{eq:spinor_action_curved}
S_\psi=\int\mathrm{d}^5x\,e\left[-\frac{1}{2}\left(\bar{\psi}e^M_{\bar{M}}\gamma^{\bar{M}}\nabla_M\psi
-\nabla_M\bar{\psi}e^M_{\bar{M}}\gamma^{\bar{M}}\psi\right)-V(\bar{\psi},\psi)\right]\,,
\end{equation}
where $e=\sqrt{-g}$ is the determinant of the vielbeins $e_M^{\bar{M}}$. A potential term with self-interactions of the spinor field could be included in the action~\eqref{eq:spinor_action_curved}. The variation of the action~\eqref{eq:spinor_action_curved} with respect to the spinor field yields
\begin{align}\label{eq:spinor_curved}
  e_{\bar{M}}^M\gamma^{\bar{M}}\nabla_M\psi+\frac{\partial V}{\partial \bar{\psi}}=0\,, \quad \nabla_M\bar{\psi}e_{\bar{M}}^M\gamma^{\bar{M}}-\frac{\partial V}{\partial \psi}=0\,.
\end{align}
While varying the action with respect to the vielbeins $e_M^{\bar{M}}$ gives the Belinfante tensor for the spinor field
\begin{align}\label{eq:EM_tensor_curved}
  T_{\bar{M}}^M=&-\frac{1}{e}\,\frac{\delta S_\psi}{\delta e_M^{\bar{M}}} \nonumber \\
  =& \frac{1}{4}\left[\bar{\psi}\left(\gamma^Me^N_{\bar{M}}+\gamma_{\bar{M}}g^{MN}\right)\nabla_N\psi
  -\nabla_N\bar{\psi}\left(\gamma^Me^N_{\bar{M}}+\gamma_{\bar{M}}g^{MN}\right)\psi\right] \nonumber \\
  &+e_{\bar{M}}^M\left[\frac{1}{2}\left(\bar{\psi}\frac{\partial V}{\partial \bar{\psi}}
  +\frac{\partial V}{\partial \psi}\psi\right)-V\right]\,.
\end{align}
The energy-momentum tensor~\eqref{eq:EM_tensor_curved} is not only conserved but also symmetric. Therefore, it can act as a source of the gravitational field.

Generally, solutions of a spinor field will be rather complicated in the presence of gravitation. Therefore, we will not exactly restrict the potential term $V(\bar{\psi},\psi)$ for the spinor field $\psi$. There are many possibilities to construct Lorentz invariant terms responsible for the potential of the spinor ﬁeld. The simplest Lorentz invariant term is $\bar{\psi}\psi$. Without loss of generality, we adopt the following potential term for the spinor field in the actions~\eqref{eq:spinor_action_flat} and~\eqref{eq:spinor_action_curved}
\begin{equation}\label{eq:potential_V}
  V(\bar{\psi},\psi)=V(\bar{\psi}\psi)\,,
\end{equation}
which is a function of the Lorentz invariant $\bar{\psi}\psi$ and generally involves a nontraditional potential. It is worth mentioning that the mass term $m\bar{\psi}\psi$ will be vanished for a massless spinor field. Before exploring solutions of the spinor field coupling with gravitation, we will begin with an illuminating solution for the spinor field in the absence of gravitation.

\section{Exact solutions in flat spacetime}
\label{sec:solutions_flat}

Solutions of a free spinor field in four-dimensional flat spacetime have been obtained in the literature. We intend to study solutions of a spinor field in five-dimensional flat spacetime, which may provide us insights into the subject. We use coordinates $x^{\mu}$ and $y$ to label the four-dimensional hypersurface and an extra spatial dimension, respectively. The five-dimensional spacetime is characterized by the background metric
\begin{equation}\label{eq:line_element_flat}
\mathrm{d}s^2=\eta_{\mu\nu}\mathrm{d}x^\mu\mathrm{d}x^\nu+\mathrm{d}y^2\,,
\end{equation}
where $\eta_{\mu\nu}$ is the four-dimensional Minkowski metric, and $y=x^5$ is the extra-dimensional coordinate. The spinor field is set to depend only on the extra spatial dimension $y$. Normally, a spinor field has four complex components in five-dimensional spacetime. For the sake of simplicity, however, we assume that the solution of a real spinor field in five-dimensional bulk spacetime has the following form~\cite{Dzhunushaliev:2011df}
\begin{equation}\label{eq:spinor}
  \psi(y)=\left(p(y),0,q(y),0\right)^{\mathrm{T}}\,,
\end{equation}
which has two nonvanishing real components $p(y)$ and $q(y)$. Here, ``$\mathrm{T}$'' denotes the transposition of a matrix. The Lorentz invariant $\bar{\psi}\psi$ for the real spinor field assumption~\eqref{eq:spinor} is given by $\bar{\psi}\psi=2p(y)q(y)$.

Substituting the real spinor field assumption~\eqref{eq:spinor} and the potential term~\eqref{eq:potential_V} into Eq.~\eqref{eq:spinor_EOM_flat}, the equations of motion for the spinor field are reduced to
\begin{align}\label{eq:pq_flat_1}
p'+\frac{\partial V}{\partial \left(\bar{\psi}\psi\right)} p=0\,, \quad
q'-\frac{\partial V}{\partial \left(\bar{\psi}\psi\right)} q=0\,,
\end{align}
where the prime denotes the derivative with respect to the extra-dimensional coordinate $y$. For the ansatz~\eqref{eq:line_element_flat} and the assumption~\eqref{eq:potential_V}, we find the following solution
\begin{align}\label{eq:pq_flat_sol}
  p(y)=c_2 \exp\left[-f(2c_1)y\right]\,, \quad q(y)=\frac{c_1}{c_2}\exp\left[f(2c_1)y\right]\,.
\end{align}
Here, $c_1$ and $c_2$ are integration constants and $f(2c_1)\equiv\frac{\partial V}{\partial \left(\bar{\psi}\psi\right)}$. It is obvious that the Lorentz invariant $\bar{\psi}\psi=2p(y)q(y)=\text{const}$. For the case of a free massive spinor field with
\begin{align}\label{eq:Vphi_flat_1}
  V(\bar{\psi}\psi)=m\bar{\psi}\psi\,,
\end{align}
the solutions of $p(y)$ and $q(y)$ can be written as
\begin{align}\label{eq:pq_flat_sol_1}
  p(y)=c_1 \exp\left(-m\, y\right)\,, \quad q(y)=c_2 \exp\left(m\, y\right)\,,
\end{align}
where $m$ is the mass parameter of the spinor field $\psi$.

One can derive a first integral from Eq.~\eqref{eq:pq_flat_1} as
\begin{align}\label{eq:first_integral_flat}
  pq=\frac{1}{2}\bar{\psi}\psi=\mathcal{C}_1 \,,
\end{align}
where $\mathcal{C}_1$ is an integration constant. It is clear that the solution~\eqref{eq:pq_flat_sol} and the simpler one~\eqref{eq:pq_flat_sol_1} satisfy the first integral~\eqref{eq:first_integral_flat}. The energy density of the spinor field with respect to a static observer $U^{\bar{M}}=(-1,0,0,0,0)$ is generally given by
\begin{align}\label{eq:ED_flat}
  \rho(y)&=T_{\bar{M}\bar{N}}U^{\bar{M}}U^{\bar{N}}=\left[p'(y)q(y)-p(y)q'(y)\right]\,.
\end{align}

Further, we will investigate exact solutions of a spinor field in the presence of gravitation.

\section{Spinor walls in warped spacetime}
\label{sec:model}

We explore solutions of a spinor field minimally coupling with gravitation in five-dimensional asymptotically AdS spacetime in which the extra spatial dimension is curved (``warped''). Such system differs from the system composed of gravitation and a scalar field. Generally speaking, a coupled system consisting of different components of the spinor field is complicated, especially in the case of the spinor field coupled to gravitation. For the sake of simplicity, a real spinor field will be taken into account. Considering the $Z_2$ symmetry of the hypersurface and the energy density distribution of the spinor field, we speculate the warp form of the hypersurface and the abstract form of the spinor field solution. Further, we will employ a reconstruction technique that the action of the spinor field is obtained according to the above speculations regarding the hypersurface and the spinor field solution. As in section~\ref{sec:solutions_flat}, we suppose that the real spinor field possesses an abstract form as Eq.~\eqref{eq:spinor}.

\subsection{The setup}

Within the context of general relativity, the bulk Einstein equation is
\begin{equation}\label{eq:GR}
  R_{\bar{M}}^M-\frac{1}{2}e_{\bar{M}}^MR=T_{\bar{M}}^M\,.
\end{equation}
In this paper, we are interested in static flat walls with the four-dimensional Poincar\'{e} symmetry, for which the general form of the metric (the general ansatz for a domain wall) is given by
\begin{equation}\label{eq:line_element_curved}
\mathrm{d}s^2=a^2(y)\eta_{\mu\nu}\mathrm{d}x^\mu\mathrm{d}x^\nu+\mathrm{d}y^2\,,
\end{equation}
where $a(y)$ is the warp factor, $\eta_{\mu\nu}$ is the four-dimensional Minkowski metric, and $y=x^5$ is the extra-dimensional coordinate. Note that the warp factor $a(y)$ and the background spinor field $\psi(y)$ are merely functions of $y$ for static flat walls. Using the metric ansatz~\eqref{eq:line_element_curved}, the spin connection $\omega_M^{\bar{A}\bar{B}}$~\eqref{eq:spin_connection} and the generator of spinor representation $S_{\bar{A}\bar{B}}$ \eqref{eq:generator} can be worked out. The nonvanishing components of $\omega_M=\frac{\mathrm{i}}{2}\omega_M^{\bar{A}\bar{B}}S_{\bar{A}\bar{B}}$ are
\begin{align}\label{eq:nonvanishing_connection}
  \omega_\mu=\frac{1}{2}\left(\partial_y a\right)\gamma_\mu\gamma_5 \,.
\end{align}

With the above metric~\eqref{eq:line_element_curved} and spin connection~\eqref{eq:nonvanishing_connection}, Eqs.~\eqref{eq:GR} and \eqref{eq:spinor_curved} are reduced to
\begin{subequations}\label{eq:1_MN}
\begin{align}
 &(\mu,\nu): \quad \frac{a'^{2}}{a^2}+\frac{a''}{a}=\frac{1}{3}\left[2\frac{\partial V}{\partial\left(\bar{\psi}\psi\right)}pq-V\right]\,, \label{eq:1_MN_munu}\\
 &(y,y): \quad \frac{a'^{2}}{a^2}=\frac{1}{6}\left[\left(p'q-pq'\right)
+2\frac{\partial V}{\partial\left(\bar{\psi}\psi\right)}pq-V\right]\,,\label{eq:1_MN_55}
\end{align}
\end{subequations}
and
\begin{subequations}\label{eq:2_pq}
\begin{align}
p'+2\frac{a'}{a}p+\frac{\partial V}{\partial\left(\bar{\psi}\psi\right)}p&=0\,,\label{eq:2_pqp} \\
q'+2\frac{a'}{a}q-\frac{\partial V}{\partial\left(\bar{\psi}\psi\right)}q&=0\,,\label{eq:2_pqq}
\end{align}
\end{subequations}
respectively. From Eq.~\eqref{eq:1_MN}, we have
\begin{align}\label{eq:1_munu55}
  \frac{a''}{a}-\frac{a'^{2}}{a^2}=\frac{1}{3}\left(pq'-p'q\right)\,.
\end{align}
From Eq.~\eqref{eq:2_pq}, the following first integral can be found
\begin{align}\label{eq:first_integral_curved}
  pqa^4=\frac{1}{2}\bar{\psi}\psi a^4= \mathcal{C}_2 \,,
\end{align}
where $\mathcal{C}_2$ is an integration constant. The first integral~\eqref{eq:first_integral_curved} is a generalization of Eq.~\eqref{eq:first_integral_flat} in the presence of gravitation. The Lorentz invariant does not depend on the form of the potential $V(\bar{\psi}\psi)$. This implies that the dynamics of the spinor field is different from that of a scalar field, especially in the presence of gravitation. In warped spacetime, the first integral~\eqref{eq:first_integral_curved} will be a significant constraint on the solutions of the spinor field.

It is worth pointing out that, however, only three equations are independent in this system in the light of the contracted Bianchi identity. However, in Eqs.~\eqref{eq:1_MN} and \eqref{eq:2_pq}, arbitrary three equations are not completely independent. Thus, we need to consider the rest equation of them. We take Eqs.~\eqref{eq:1_MN_munu}, \eqref{eq:2_pqp}, and \eqref{eq:2_pqq} as the beginning of solving this system. We will see that the solutions of Eqs.~\eqref{eq:1_MN_munu}, \eqref{eq:2_pqp}, and \eqref{eq:2_pqq} have similar characteristics as the solutions in flat spacetime. Actually, one can also obtain a set of solutions by Eqs.~\eqref{eq:2_pq} and \eqref{eq:1_munu55}. However, these solutions are equivalent to the ones which we will present below.

Since this system satisfies the equations of motion, the energy density could be of the general form
\begin{align}\label{eq:ED_curved}
  \rho(y)&=T_{MN}U^MU^N=\left(R_{MN}-\frac{1}{2}g_{MN}R\right)U^MU^N\,.
\end{align}
For a static observer $U^{M}=(-1/a(y),0,0,0,0)$ lying on the wall, the distribution of the observed energy density along extra dimension $y$ would be
\begin{align}\label{eq:ED_curved_trans}
  \rho(y)=-3\left(\frac{a'^2}{a^2}+\frac{a''}{a}\right)\,.
\end{align}
It is worth noting that in this energy density, the contribution of the cosmological constant need to be subtracted, so that the minimum energy density (if it exists) vanishes.

There are two significant parameters for domain walls with $Z_2$ symmetry~\cite{Vilenkin:1985v}, which will be employed in the following discussions.
\begin{itemize}
  \item The first one is the tension of the wall, which measures the energy per unit volume on the wall. The tension of the wall is defined as the integral of the energy density along the coordinate $y$
      \begin{align}\label{eq:walltension}
        \varepsilon=\int_{\Sigma}\sqrt{-g}\rho(y)\mathrm{d}y<\infty\,,
      \end{align}
      where $g$ is the determinant of the metric and $\Sigma$ is the integration interval of the fifth dimension $y$. For an infinite extra dimension, the integration interval will be $\left(-\infty,\infty\right)$. Generally speaking, if the tension of the wall is finite, the solution of the wall is regular.
  \item The second one is the thickness of the wall, which evaluates the typical length scale of the scale variation of the wall. The thickness of the wall in the domain wall case is defined as the reciprocal of the factor $k$ appearing in front of the $y$. We denote the thickness of the wall by $\sigma=1/k$ and the above definition leads to
      \begin{align}\label{eq:wallthickness}
        \int_{-\frac{\sigma}{2}}^{\frac{\sigma}{2}}\sqrt{-g}\rho(y)\mathrm{d}y\approx c\times \varepsilon\,,
      \end{align}
      where $c$ ranges from $0\%$ to $100\%$. For the following $Z_2$ domain wall solution in flat spacetime
      \begin{align}
        \phi(y)=v_0 \tanh(k y)\,,
      \end{align}
      $c$ is approximately $64\%$. Here, $k$ is a parameter and the length scale $1/k$ is related to the thickness of the wall.
\end{itemize}

By means of the curvature of the bulk, the bulk spacetime could be classified into two kinds: the constant and nonconstant curvature spacetimes. We will discuss solutions in these two kinds of spacetimes in the following two subsections.

\subsection{Nonconstant curvature solutions}

\begin{figure}[htb]
\centering
\subfloat[\ Warp factor $a(\tilde y)$ \label{subfig:ay1}]{
\includegraphics[width=2.6in]{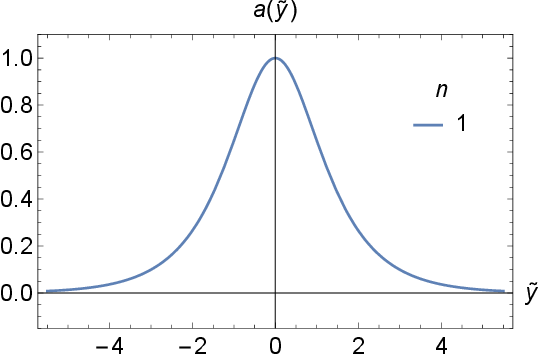}}
\hspace{0.5in}
\subfloat[\ Energy density $\tilde{\rho}\left(\tilde{y}\right)$ \label{subfig:rhoy1}]{
\includegraphics[width=2.6in]{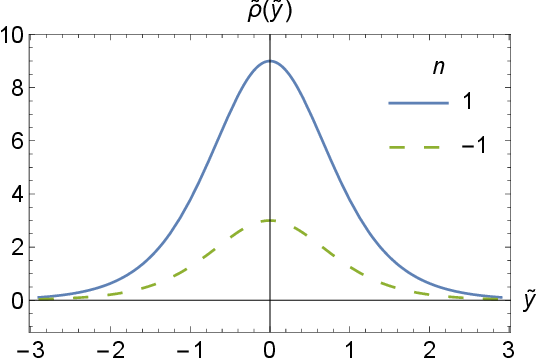}}
\hspace{0.5in}
\subfloat[\ Warp factor $a(\tilde y)$ \label{subfig:ay1_}]{
\includegraphics[width=2.6in]{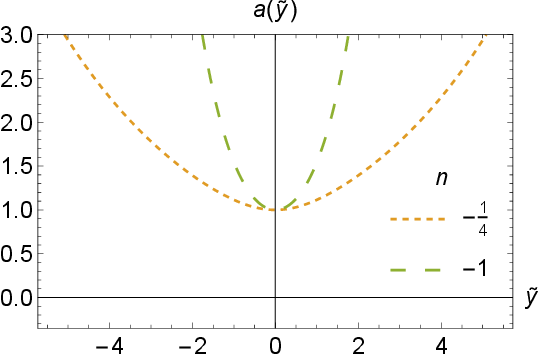}}
\hspace{0.5in}
\subfloat[\ Energy density $\tilde{\rho}\left(\tilde{y}\right)$ \label{subfig:rhoy1_}]{
\includegraphics[width=2.6in]{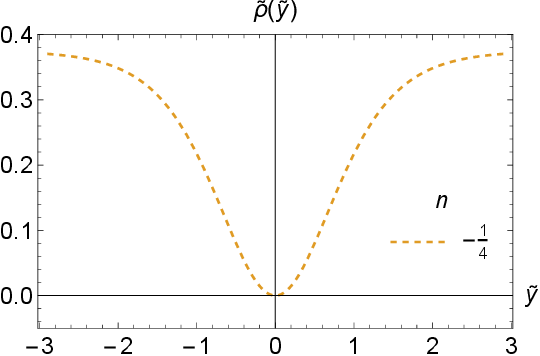}}
\caption{Plots of $a(\tilde y)$ and $\tilde{\rho}\left(\tilde{y}\right)$, where $a(\tilde y)$ is the warp factor in Eq.~\eqref{eq:warp_factor1} and $\rho(\tilde y)$ is the energy density in Eq.~\eqref{eq:ED_curved_trans}. The parameter $n$ is set to $1$, $-\frac{1}{4}$, and $-1$. The warp factors (left panels) with a blue solid line, orange dashed line, and green dashing line correspond to the energy densities (right panels) with a blue solid line, orange dashed line, and green dashing line, respectively.}\label{fig:ay&rhoy1}
\end{figure}

We will investigate exact solutions of a spinor field in warped spacetime. Firstly, we begin with a given bulk spacetime which is characterized by Eq.~\eqref{eq:line_element_curved} with the warp factor
\begin{align}\label{eq:warp_factor1}
 a(y)=\mathrm{sech}^n(k y)\,.
\end{align}
An obvious property is that the warp factor~\eqref{eq:warp_factor1} is smooth rather than singular at the location of $y=0$. Such property indicates that the curvature singularity at the location of the codimension one hypersurface will be eliminated. The profile of the warp factor~\eqref{eq:warp_factor1} for positive $n$ is shown in Fig.~\ref{subfig:ay1}. The bulk spacetime with the warp factor~\eqref{eq:warp_factor1} with nonzero $n$ is asymptotically AdS at the boundary of the extra dimension. Generally speaking, the $n$ in this warp factor~\eqref{eq:warp_factor1} could be negative, zero, and positive. As we will demonstrate in section~\ref{sec:localization}, if $n$ is positive, a localized graviton zero mode can be obtained. If $n=0$, the bulk spacetime is flat. Nevertheless, we also discuss the case of negative $n$, because it has similar behaviour to the Janus solution~\cite{Bak:2003bgh}, one of the early examples on the study of defects and interfaces in field theories.

By solving the equations of motion, we find a solution of spinor domain wall generated by the spinor field~\eqref{eq:spinor} with the following components $p(y)$ and $q(y)$
\begin{subequations}\label{eq:pq_curved_sol1}
\begin{align}
  p(y)&=c_3 \exp\left[-\frac{c_1}{2 c_2}y+D(y)\right] \cosh ^{2 n}(k y)\,, \\
  q(y)&=\frac{c_2}{c_3}\exp\left[\frac{c_1}{2 c_2}y-D(y)\right] \cosh ^{2 n}(k y)\,,
\end{align}
\end{subequations}
and the potential
\begin{equation}\label{eq:Vphi_curved_1}
 V(\bar{\psi}\psi)= \frac{c_1}{2 c_2}\bar{\psi}\psi-
 6 k^2 n^2\left[1-\left(\frac{\bar{\psi}\psi}{2 c_2}\right)^{-\frac{1}{2 n}}\right]\,,
\end{equation}
where we have defined a new function $D(y)$ as
\begin{align}\label{eq:function_D}
  D(y)=\mathrm{i}\frac{3kn}{4c_2}\mathrm{sgn}(ky)\,
  \mathrm{B}_{\cosh^2(ky)}\left(-2n-\frac{1}{2},\frac{1}{2}\right)\,,
\end{align}
and $\frac{c_1}{2 c_2}$ is identified as the mass parameter (we can also see it from Eq.~\eqref{eq:pq_flat_sol_1}). Here, $\mathrm{sgn}$ is the sign function and $\mathrm{B}$ is the incomplete beta function. According to the assumption~\eqref{eq:spinor}, the function $D(y)$~\eqref{eq:function_D} should be real. Due to this incomplete beta function $\mathrm{B}_{\cosh^2(ky)}\left(-2n-\frac{1}{2},\frac{1}{2}\right)$ is a complex function, we need to find the conditions that the function $D(y)$~\eqref{eq:function_D} is real. If we take $n$ to be zero, positive integers, and positive half-integers, the function $D(y)$ is real. In addition to these values, the function $D(y)$ is complex. In this case, the imaginary component should be discarded because it is introduced in the process of solving the equations, where the variable $y$ has been analytically continued to the complex domain. From the perspective of the equations of motion, the resulting solution of a real differential equation will be real. In essence, the new defined function $D(y)$ can always be a real function, owing to the real component of the incomplete beta function $\mathrm{B}$ can be canceled by using the integration constant $c_3$. For the potential term~\eqref{eq:Vphi_curved_1}, this spinor field has a nonlinear potential as a correction to the linear term $\bar{\psi}\psi$. The additional constant in the potential~\eqref{eq:Vphi_curved_1} could be accounted for the effective cosmological constant. In fact, the equations of motion~\eqref{eq:1_MN} and \eqref{eq:2_pq} require that the integration constant $c_1$ is zero. For comparison with the solutions in flat spacetime, we have retained the integration constant $c_1$ so that $c_1=0$ is a special case.

To demonstrate the behaviour of the warp factors and the corresponding energy densities for the following spinor field solutions, we will henceforth introduce the dimensionless quantities $\tilde{y}=k y$ and $\tilde{\rho}\left(\tilde{y}\right)=\rho(\tilde y)/k^2$ in the figures below to display their profiles.

For the warp factor~\eqref{eq:warp_factor1}, the energy density is calculated as
\begin{align}\label{eq:ed1}
  \rho(y)=3 k^2 n (2 n+1) \text{sech}^2(ky)\,.
\end{align}
If $n$ takes $0$ or $-\frac{1}{2}$, the energy density will be zero. The warp factor~\eqref{eq:warp_factor1} and the corresponding energy density of the spinor field are depicted in Fig.~\ref{fig:ay&rhoy1}. Note that, we have deducted the contribution of the effective cosmological constant so that the minimum value of the energy density is zero. As shown in Fig.~\ref{fig:ay&rhoy1}, for $-\frac{1}{2}<n<0$, the energy density increases with $|y|$ and approaches to a constant as $\tilde{y}\rightarrow\pm\infty$. There is a trap of the energy density at $\tilde{y}=0$. For $n<-\frac{1}{2}$ and $n>0$, the energy density decreases with $|y|$ down and approaches to zero as $\tilde{y}\rightarrow\pm\infty$. The energy density of the spinor field centers on the vicinity of the hypersurface. The more concentrated the energy density of the spinor field is, the more warped the extra dimension is. There is a wall between two sides of the hypersurface. The wall is similar to a domain wall, a codimension one soliton generated by scalar fields. This solution of the spinor field can be viewed as a thick domain wall interpolating between two asymptotically AdS vacua.

In the case of the warp factor~\eqref{eq:warp_factor1}, according to~\eqref{eq:walltension}, we obtain the tension of the wall in the domain $(-\infty,\infty)$ for $n>0$
\begin{align}\label{eq:tension11}
  \varepsilon=3 \sqrt{\pi } k n \frac{\Gamma (2 n+2)}{\Gamma \left(2 n+\frac{3}{2}\right)}\,,
\end{align}
where $\Gamma$ is the Euler gamma function. The tension of the trap in the domain $(-\frac{\sigma}{2},\frac{\sigma}{2})$ for $-\frac{1}{2}<n<0$ is
\begin{align}\label{eq:tension12}
  \varepsilon= \frac{3 k}{2}\tanh \left(\frac{k \sigma }{2}\right) \text{sech}^{4 n}\left(\frac{k \sigma }{2}\right) \left[2 n+1-\, _2F_1\left(1,-2 n;2 (n+1);-e^{k \sigma }\right)\right]\,,
\end{align}
where the special function $_2F_1$ is the hypergeometric function. The tension of the wall in the domain $(-\frac{\sigma}{2},\frac{\sigma}{2})$ for $n<-\frac{1}{2}$ is
\begin{align}\label{eq:tension13}
  \varepsilon= &3 \sqrt{\pi } \sqrt{k^2} n \frac{\Gamma (2 n+2)}{\Gamma \left(2 n+\frac{3}{2}\right)}-3 k n \text{sgn}\left(\frac{k \sigma }{2}\right) \text{sech}^{4 n+2}\left(\frac{k \sigma }{2}\right) \nonumber\\ &\times\, _2F_1\left(\frac{1}{2},2 n+1;2 (n+1);\text{sech}^2\left(\frac{k \sigma }{2}\right)\right)\,.
\end{align}
The integral~\eqref{eq:walltension} is divergent for $n<-\frac{1}{2}$ and $-\frac{1}{2}<n<0$, and is convergent for $n>0$.
Consequently, the solution~\eqref{eq:pq_curved_sol1} and ~\eqref{eq:Vphi_curved_1} is globally regular for $n>0$ and is locally regular for $-\frac{1}{2}<n<0$ and $n<-\frac{1}{2}$.

For the case where the bulk field is a scalar field, the profile functions of the Jauns solution~\cite{Freedman:2004fnss} typically are
\begin{align}
&a(y)=\frac{1}{2}\left[1+\sqrt{1-2 \gamma^{2}} \cosh(2ky)\right]\,, \label{eq:Janus} \\
&\phi(y)=\phi_{0}+\frac{1}{\sqrt{2}} \log \left[\frac{1+\sqrt{1-2 \gamma^{2}}+\sqrt{2} \gamma \tanh(ky)}{1+\sqrt{1-2 \gamma^{2}}-\sqrt{2} \gamma \tanh(ky)}\right]\,.
\end{align}
Here, the parameter $\gamma \in\left[0, \frac{1}{\sqrt{2}}\right]$. In the Janus solution, the scalar field solution only varying spatially resembles a domain wall which is a typical soliton configuration. There is a possibility of a wall-like configuration for the spinor field. The warp factor~\eqref{eq:Janus} is also smooth at the location of $y=0$ and the curvature singularity at the location of the codimension one hypersurface will be eliminated. A common property of the warp factors~\eqref{eq:warp_factor1} and \eqref{eq:Janus} is that the bulk spacetime is asymptotically AdS at the boundary $\tilde{y}\rightarrow\infty$.

\begin{figure}[htb]
\centering
\subfloat[\ Warp factor $a(\tilde y)$ \label{subfig:ayJ}]{
\includegraphics[width=2.6in]{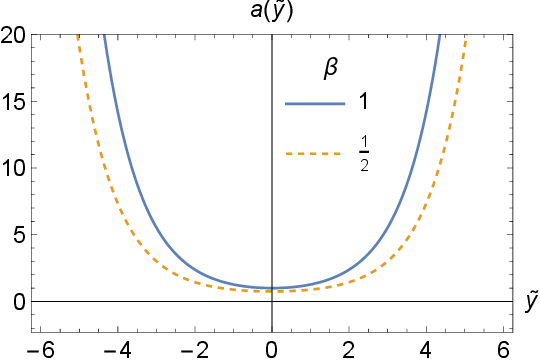}}
\hspace{0.5in}
\subfloat[\ Energy density $\tilde{\rho}\left(\tilde{y}\right)$ \label{subfig:rhoyJ}]{
\includegraphics[width=2.6in]{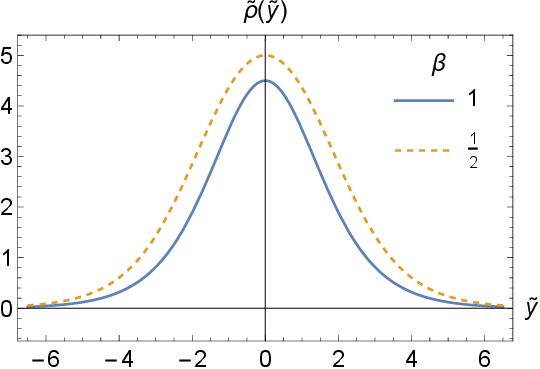}}
\caption{Plots of $a(\tilde y)$ and $\tilde{\rho}\left(\tilde{y}\right)$, where $a(\tilde y)$ is the warp factor in Eq.~\eqref{eq:warp_factor_Janus} and $\rho(\tilde y)$ is the energy density in Eq.~\eqref{eq:ED_curved_trans}. The parameter $\beta$ is set to $1$ and $\frac{1}{2}$.}\label{fig:ay&rhoyJ}
\end{figure}

As a deformation of the warp factor~\eqref{eq:warp_factor1}, the Janus solution~\eqref{eq:Janus} is rather difficult to solve the equations of motion~\eqref{eq:1_MN} and \eqref{eq:2_pq}. For the convenience of solving the equations of motion, we take $\beta\equiv\sqrt{1-2\gamma^{2}}, \beta\in\left[0,1\right]$ and the $\cosh(2ky)$ is replaced by the $\cosh(ky)$ in Eq.~\eqref{eq:Janus}. Therefore, we begin with the warp factor of the form
\begin{align}\label{eq:warp_factor_Janus}
  a(y)=\frac{1}{2}\left[1+\beta \cosh(ky)\right]\,.
\end{align}
One can obtain the $p(y)$ and $q(y)$ of Eq.~\eqref{eq:spinor} as follows
\begin{subequations}\label{eq:pq_Janus}
\begin{align}
  p(y)&=c_3 \exp[-\frac{c_1}{2 c_2}y-E(y)] \frac{1}{[1+\beta  \cosh \left(k y)\right]^2} \,, \\
  q(y)&=\frac{c_2}{c_3}  \exp[\frac{c_1}{2 c_2}y+E(y)] \frac{1}{\left[1+\beta  \cosh (k y)\right]^2} \,,
\end{align}
\end{subequations}
and the potential of the spinor field
\begin{align}\label{eq:Vphi_curved_Janus}
  V(\bar{\psi}\psi)= \frac{c_1}{2 c_2}\bar{\psi}\psi- 6 k^2 \left[1+\frac{\beta^2 \left(\beta^2+4\right)}{8}\left(\frac{\bar{\psi}\psi }{2 c_2}\right)+ \left(1-\beta^2\right) \left(\frac{\bar{\psi}\psi }{2 c_2}\right)^{\frac{1}{2}}+ 2 \left(\frac{\bar{\psi}\psi }{2 c_2}\right)^{\frac{1}{4}} \right]\,,
\end{align}
where the function $E(y)$ is given by
\begin{align}
  E(y)=&\frac{1}{8 c_2}\left\{3 k^2 \beta^2\left(4+\beta^2\right) y \right. \nonumber \\ & \left. +k \left[\beta \left(12 +33 \beta^2\right)\sinh(k y) + 3\beta^2 \left(2+\beta^2\right) \sinh(2 k y)+\beta^3 \sinh(3 k y)\right] \right\} \,.
\end{align}
Here, $\frac{c_1}{2 c_2}$ could also be viewed as the mass of the spinor field. In the potential~\eqref{eq:Vphi_curved_Janus}, there are nonlinear terms as a correction and an extra constant term could be accounted for the effective cosmological constant. It is worth stressing that according to the requirement of the equations of motion, the integration constant satisfies $c_1=\frac{3}{4} k^2\beta ^2 \left(4+\beta ^2\right)$. Nevertheless, we still retain $c_1$ as an arbitrary constant to compare with the solutions in flat spacetime. We depict the warp factor~\eqref{eq:warp_factor_Janus} and the corresponding energy density in Fig.~\ref{fig:ay&rhoyJ}. For the warp factor~\eqref{eq:warp_factor_Janus} which is similar to the warp factor~\eqref{eq:warp_factor1} with $n=-1$, the corresponding solution is not globally regular.

From the aforementioned results, one can find that there is no singularity in the geometry of the bulk. This resembles a thick domain wall. A remarkable property is that this solution has a holographic interpretation in terms of the theory of living on the boundary~\cite{Bak:2003bgh}.

\begin{figure}[htb]
\centering
\subfloat[\ Warp factor $a(\tilde y)$ \label{subfig:ay2}]{
\includegraphics[width=2.6in]{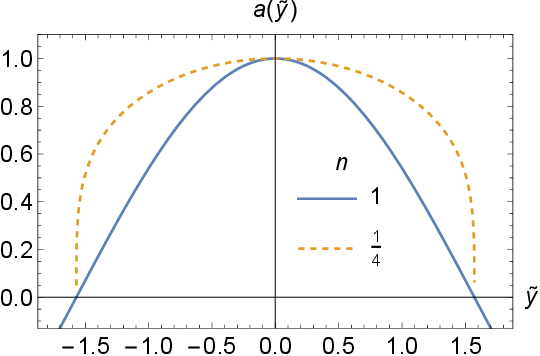}}
\hspace{0.5in}
\subfloat[\ Energy density $\tilde{\rho}\left(\tilde{y}\right)$ \label{subfig:rhoy2}]{
\includegraphics[width=2.6in]{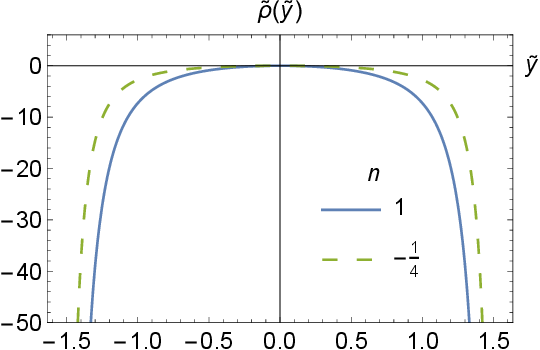}}
\hspace{0.5in}
\subfloat[\ Warp factor $a(\tilde y)$ \label{subfig:ay2_}]{
\includegraphics[width=2.6in]{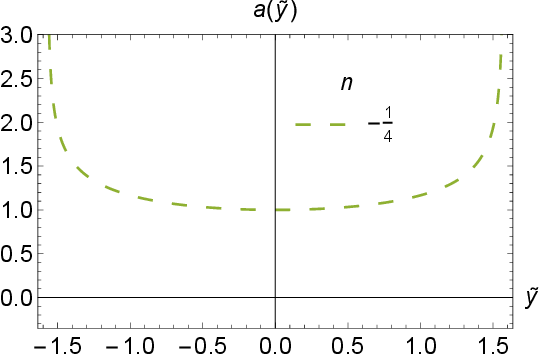}}
\hspace{0.5in}
\subfloat[\ Energy density $\tilde{\rho}\left(\tilde{y}\right)$ \label{subfig:rhoy2_}]{
\includegraphics[width=2.6in]{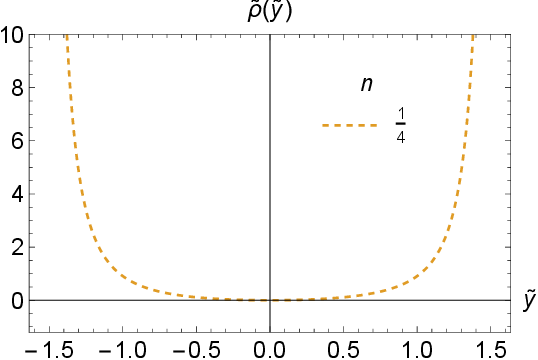}}
\caption{Plots of $a(\tilde y)$ and $\tilde{\rho}\left(\tilde{y}\right)$, where $a(\tilde y)$ is the warp factor in Eq.~\eqref{eq:warp_factor2} and $\rho(\tilde y)$ is energy density in Eq.~\eqref{eq:ED_curved_trans}. The parameter $n$ is set to $1$, $\frac{1}{4}$, and $-\frac{1}{4}$. The warp factors (left panels) with a blue solid line, orange dashed line, and green dashing line correspond to the energy densities (right panels) with a blue solid line, orange dashed line, and green dashing line, respectively.}\label{fig:ay&rhoy2}
\end{figure}

The second analytic solution of the spinor field in warped spacetime will be exhibited in the following context. We choose the following warp factor~\cite{Gremm:2000gt}
\begin{equation}\label{eq:warp_factor2}
  a(y)=\cos^n(k y)\,.
\end{equation}
The warp factor is smooth at the position of the hypersurface as well. The profile of the warp factor is shown in Fig.~\ref{fig:ay&rhoy2}. This indicates that the curvature singularity is eliminated at the location of the codimension one hypersurface. For the warp factor~\eqref{eq:warp_factor2} with negative $n$, the bulk spacetime will be singular at the boundary $\tilde{y}=\pm\frac{\pi}{2}$. Therefore, the range of $\tilde{y}$ is from $-\frac{\pi}{2}$ to $\frac{\pi}{2}$. The case $n=0$ corresponds to a flat bulk spacetime. We find a solution of the spinor field~\eqref{eq:spinor} with the following $p(y)$ and $q(y)$
\begin{subequations}\label{eq:pq_curved_sol2}
\begin{align}
  p(y)&=c_3 \exp\left[-\frac{c_1}{2 c_2}y-F(y)\right] \cos^{-2 n}(k y)\,, \\
  q(y)&=\frac{c_2}{c_3}\exp\left[\frac{c_1}{2 c_2}y+F(y)\right] \cos^{-2 n}(k y)\,,
\end{align}
\end{subequations}
and the potential is determined by
\begin{equation}\label{eq:Vphi_curved_2}
 V(\bar{\psi}\psi)=\frac{c_1}{2 c_2}\bar{\psi}\psi+
 6 k^2 n^2 \left[1-\left(\frac{\bar{\psi}\psi}{2 c_2}\right)^{\frac{1}{2 n}}\right]\,,
\end{equation}
where the function $F(y)$ is given by
\begin{equation}\label{eq:function_E}
  F(y)=\frac{3 k n}{4 c_2}\mathrm{sgn}(k y)\,
  \mathrm{B}_{\cos ^2(k y)}\left(2 n-\frac{1}{2},\frac{1}{2}\right)\,,
\end{equation}
and $\frac{c_1}{2 c_2}$ is viewed as the mass. In this case, the incomplete beta function $\mathrm{B}_{\cos ^2(k y)}\left(2 n-\frac{1}{2},\frac{1}{2}\right)$ is real. Therefore, the function $F(y)$~\eqref{eq:function_E} is a real function which satisfies the assumption~\eqref{eq:spinor}. Similar to the first solution, the equations of motion~\eqref{eq:1_MN} and \eqref{eq:2_pq} require that the integration constant $c_1$ is zero. For comparison with the solutions in flat spacetime, we have retained the integration constant $c_1$ so that $c_1=0$ is a special case.

In the case of the warp factor~\eqref{eq:warp_factor2}, the energy density is computed as
\begin{align}\label{eq:ed2}
  \rho(y)=3 k^2 n (1-2 n) \tan ^2(x)\,.
\end{align}
If $n$ takes $0$ or $\frac{1}{2}$, the energy density will be a constant. The warp factor and the energy density of the spinor field are plotted in Fig.~\ref{fig:ay&rhoy2}. The energy density at the $\tilde{y}$ direction boundaries $-\frac{\pi}{2}$ and $\frac{\pi}{2}$ is positive infinity for $0<n<\frac{1}{2}$ and is negative infinity for $n<0$ and $n>\frac{1}{2}$. Nevertheless, from the local perspective, these configurations can be seen as a local wall for $n<0$ and $n>\frac{1}{2}$ or a local trap for $0<n<\frac{1}{2}$.

In line with the definition~\eqref{eq:walltension}, the tension of the wall in the domain $(-\sigma/2,\sigma/2)$ for $n<0$ and $n>\frac{1}{2}$ is given by
\begin{align}\label{eq:tension21}
  \varepsilon=& 3 k n (2 n-1) \text{sgn}\left(\frac{k \sigma }{2}\right) \nonumber\\ &\times \left[\sec ^2\left(\frac{k \sigma }{2}\right) \mathrm{B}_{\cos ^2\left(\frac{k \sigma }{2}\right)}\left(2 n-\frac{1}{2},\frac{3}{2}\right)-\tan ^2\left(\frac{k \sigma }{2}\right) \mathrm{B}_{\cos ^2\left(\frac{k \sigma }{2}\right)}\left(2 n-\frac{1}{2},\frac{1}{2}\right)\right]\nonumber\\ &
  -\frac{3 \sqrt{\pi} \sqrt{k^2}}{4}\left[\sec ^2\left(\frac{k \sigma }{2}\right)-4 n \tan ^2\left(\frac{k \sigma }{2}\right)\right]\frac{\Gamma \left(2 n-\frac{1}{2}\right)}{\Gamma (2 n-1)}\,.
\end{align}
The tension of the trap in the domain $(-\sigma/2,\sigma/2)$ for $0<n<\frac{1}{2}$ is
\begin{align}\label{eq:tension22}
  \varepsilon=& 3 k n (2 n-1) \text{sgn}(k \sigma ) \mathrm{B}_{\cos ^2\left(\frac{k \sigma }{2}\right)}\left(2 n-\frac{1}{2},\frac{3}{2}\right)\,.
\end{align}
This result implies that the wall or trap is locally regular. If the thickness of the wall is extended to the boundary of the extra dimension $\sigma=\pi$, the wall will be singular. The integral~\eqref{eq:walltension} is divergent for any $n$, because the energy density at the $\tilde{y}$ direction boundaries $-\frac{\pi}{2}$ and $\frac{\pi}{2}$ is positive infinity for $0<n<\frac{1}{2}$ and is negative infinity for $n<0$ and $n>\frac{1}{2}$. Nevertheless, from the local perspective, the integral~\eqref{eq:walltension} is locally convergent and regular. Therefore, these configurations can be seen as a local wall for $n<0$ and $n>\frac{1}{2}$ or a local trap for $0<n<\frac{1}{2}$.

We will show the third solution which displays an analogous behaviour of a warped bulk compared with the first solution. We choose the warp factor as follows
\begin{equation}\label{eq:warp_factor3}
  a(y)=\exp\left[-(k y)^2\right]\,.
\end{equation}
The warp factor possesses a simplified form of the normal distribution and is smooth at the position of the hypersurface. Therefore, there do not exist singularities in the bulk. In this warped spacetime, a spinor domain wall is generated by the spinor field~\eqref{eq:spinor} with the following $p(y)$ and $q(y)$
\begin{subequations}\label{eq:pq_curved_sol3}
\begin{align}
  p(y)&=c_3 \exp\left[-\frac{c_1}{2 c_2}y+G(y)\right] \exp\left[2 (k y)^2\right] \,, \\
  q(y)&=\frac{c_2}{c_3}\exp\left[\frac{c_1}{2 c_2}y-G(y)\right] \exp\left[2 (k y)^2\right] \,,
\end{align}
\end{subequations}
and the potential is determined by
\begin{equation}\label{eq:Vphi_curved_3}
 V(\bar{\psi}\psi)=\frac{c_1}{2 c_2}\bar{\psi}\psi-6k^2\ln \left(\frac{\bar{\psi}\psi}{2 c_2}\right)\,,
\end{equation}
where the function $G(y)$ is given by
\begin{equation}\label{eq:function_F}
  G(y)=\frac{3 \sqrt{\pi} k}{4 c_2}\mathrm{erf}\left(2 k y\right) \,,
\end{equation}
with $\mathrm{erf}$ the error function. Obviously, the function~\eqref{eq:function_F} is a real function in the real domain. Similar to previous solutions, when $c_1$ is zero, this solution satisfies the equations of motion~\eqref{eq:1_MN} and \eqref{eq:2_pq}.

\begin{figure}[htb]
\centering
\subfloat[\ Warp factor $a(\tilde y)$ \label{subfig:ay3}]{
\includegraphics[width=2.6in]{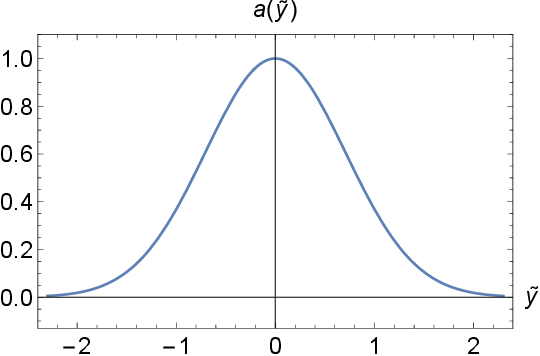}}
\hspace{0.5in}
\subfloat[\ Energy density $\tilde{\rho}\left(\tilde{y}\right)$ \label{subfig:rhoy3}]{
\includegraphics[width=2.6in]{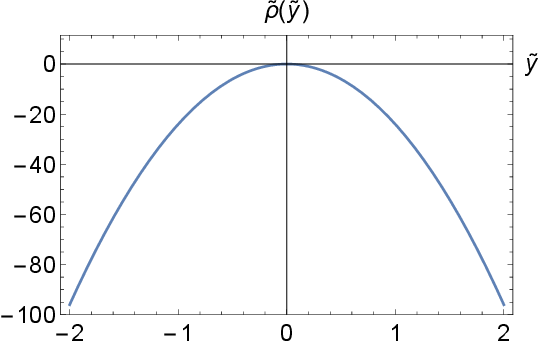}}
\caption{Plots of $a(\tilde y)$ and $\tilde{\rho}\left(\tilde{y}\right)$, where $a(\tilde y)$ is the warp factor in Eq.~\eqref{eq:warp_factor3} and $\rho(\tilde y)$ is the energy density in Eq.~\eqref{eq:ED_curved_trans}.}\label{fig:ay&rhoy3}
\end{figure}

Substituting the warp factor into Eq.~\eqref{eq:ED_curved_trans}, one obtains the energy density as follows
\begin{align}\label{eq:ed3}
  \rho(y)=-24 k^4 y^2\,.
\end{align}
The warp factor and the energy density of the spinor field are depicted in Fig.~\ref{fig:ay&rhoy3}. As shown in Fig.~\ref{fig:ay&rhoy3}, the energy density approaches to the negative infinity as $\tilde{y}\rightarrow\pm\infty$. From the local perspective, there is a local wall at the position $\tilde{y}=0$, because the boundary of the bulk along $\tilde{y}$ can not be reached physically. It is worth mentioning that we always set the minimum value of the energy density of the local wall as zero so that the vacuum energy density is zero and an effective cosmological constant is introduced.

In accordance with~\eqref{eq:walltension}, the tension of the wall in the domain $(-\sigma/2,\sigma/2)$ is
\begin{align}\label{eq:tension3}
  \varepsilon=\frac{3}{2} k \left[\sqrt{\pi } \left(2 k^2 \sigma ^2-1\right) \text{erf}(k \sigma )+2 k \sigma  e^{-k^2 \sigma ^2}\right]\,.
\end{align}
Such integral~\eqref{eq:walltension} of the energy density does not converge in the domain $(-\infty,\infty)$. It implies that the wall is locally regular. If the thickness of the wall is extended to the boundary of the extra dimension $\sigma=\infty$, the wall will be singular.

We have investigated spinor wall solutions in the bulk with nonconstant curvature. Among these solutions, constant curvature cases have been included. In the following subsection, we will specifically discuss them. All these solutions definitely satisfy the first integral~\eqref{eq:first_integral_curved}. As mentioned above, we retain the integration constant $c_1$ so that $c_1=0$ ($c_1=\frac{3}{4} k^2\beta^2 \left(4+\beta^2\right)$ for the Janus solution above) is a special case which determinately satisfies the equations of motion~\eqref{eq:1_MN} and \eqref{eq:2_pq}. The solutions in curved spacetime can be regarded as a correction to the ones in flat spacetime. Yet from the perspective of the solutions with $c_1=0$ ($c_1=\frac{3}{4} k^2\beta^2 \left(4+\beta^2\right)$ for the Janus solution above), the solutions in curved spacetime are rather different from the ones in flat spacetime since the mass term in these potentials of the spinor field actually is vanishing. Therefore, the solutions in curved spacetime could also not be considered as a correction on the ones in flat spacetime.

It is worth noting that the three sets of solutions of the spinor field are static. All the solutions of the spinor field can lead to a wall-like configuration. The wall solutions with the warp factors~\eqref{eq:warp_factor1} (including the deformation \eqref{eq:warp_factor_Janus}) and \eqref{eq:warp_factor2} are smooth at the location of the hypersurface and have a well-defined thin wall limit. The wall solution with the warp factor~\eqref{eq:warp_factor3} is also smooth at the location of the hypersurface but has a definite thickness. Moreover, in the case of the warp factor~\eqref{eq:warp_factor1}, its energy density mainly localizes in the vicinity of the hypersurface even though the extra dimension is infinite. In the case of the warp factors~\eqref{eq:warp_factor2} and \eqref{eq:warp_factor3}, their energy densities are negative infinity at boundaries and there are local walls from the local perspective.

Whether these soliton-like walls are stable or not is still crucial. Localization of gravitation is also worth investigating since it relates to effective gravitational theories on these walls. These issues will be addressed in the following sections.

\subsection{Constant curvature solutions}

Having exhibited three sets of solutions for the spinor field, we intend to discuss special solutions which correspond to three types of bulk spacetimes with constant curvature, namely, zero scalar curvature, de Sitter (dS), and AdS bulk spacetimes. One can verify that the three types of five-dimensional bulk spacetimes have three kinds of metrics characterized by Eq.~\eqref{eq:line_element_curved} with analytic warp factors~\cite{Afonso:2007abmp}, which are listed in Tab.~\ref{tab:warp}.
\begin{table}[htb]
\centering
{\renewcommand\arraystretch{1.4}
\begin{tabular}{|c|c|c|}
  \hline
  Warp factors $a(y)$ & Scalar curvature $R(y)$ & Bulk spacetimes \\
  \hline
  $\left(\frac{5}{2}ky\right)^{\frac{2}{5}}$ & $R(y)=0$ & Zero scalar curvature \\
  \hline
  $\cos^{\frac{2}{5}}\left(\frac{5}{2}ky\right)$ & $R(y)=20k^2$ & dS \\
  \hline
  $\cosh^{\frac{2}{5}}\left(\frac{5}{2}ky\right)$ & $R(y)=-20k^2$ & AdS \\
  \hline
\end{tabular}}
\caption{Table for three background spacetimes and their corresponding warp factors.}\label{tab:warp}
\end{table}

\begin{figure}[htb]
\centering
\subfloat[\ $a(\tilde y)$ for three spacetimes \label{subfig:ay_curved}]{
\includegraphics[width=2.6in]{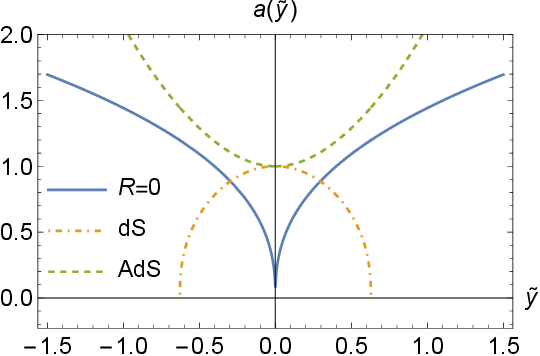}}
\hspace{0.5in}
\subfloat[\ $\tilde{\rho}\left(\tilde{y}\right)$ for the bulk spacetime with zero scalar curvature \label{subfig:rhoy_m_curved}]{
\includegraphics[width=2.6in]{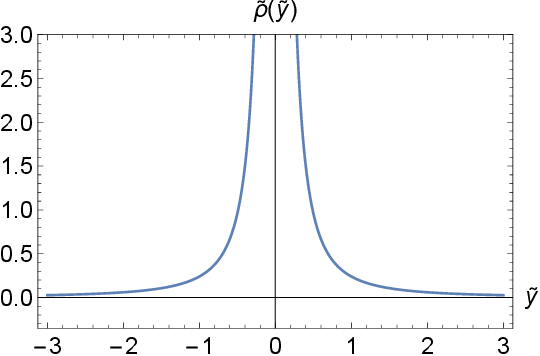}}
\hspace{0.5in}
\subfloat[\ $\tilde{\rho}\left(\tilde{y}\right)$ for the dS bulk spacetime \label{subfig:rhoy_dS_curved}]{
\includegraphics[width=2.6in]{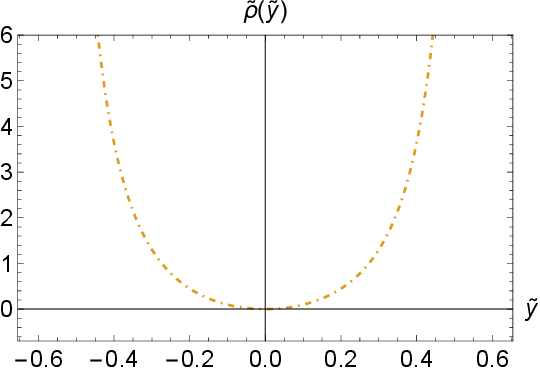}}
\hspace{0.5in}
\subfloat[\ $\tilde{\rho}\left(\tilde{y}\right)$ for the AdS bulk spacetime \label{subfig:rhoy_AdS_curved}]{
\includegraphics[width=2.6in]{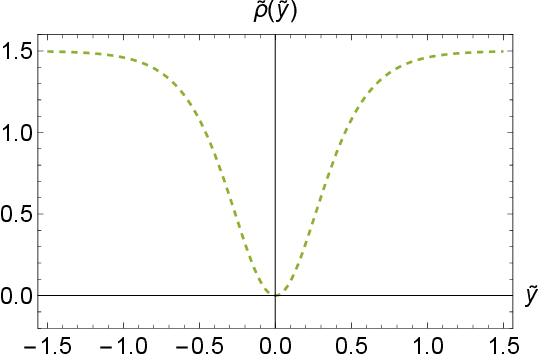}}
\caption{Plots of the warp factor $a(\tilde y)$ and the corresponding dimensionless energy density $\tilde{\rho}\left(\tilde{y}\right)$, where $a(\tilde y)$ is the warp factors in Tab.~\ref{tab:warp} and $\rho(\tilde y)$ is the energy density in Eq.~\eqref{eq:ED_curved_trans}. The warp factors (the upper left panel) with a blue solid line, orange dot-dashed line, and green dashed line correspond to the energy densities (the rest panels) with a blue solid line, orange dot-dashed line, and green dashed line, respectively.}\label{fig:ay&rhoy_curved}
\end{figure}

We plot their profiles of the warp factors in Fig.~\ref{subfig:ay_curved}. The warp factors $\cos^{\frac{2}{5}}\left(\frac{5}{2}ky\right)$ and $\cosh^{\frac{2}{5}}\left(\frac{5}{2}ky\right)$ are smooth in their domain of definition. The warp factor $\cos^{\frac{2}{5}}\left(\frac{5}{2}ky\right)$ is only well-defined in the interval $\tilde{y}\in\left[-\frac{\pi}{5},\frac{\pi}{5}\right]$. The warp factor $\left(\frac{5}{2}ky\right)^{\frac{2}{5}}$ has an inflection point at $\tilde{y}=0$ and there is a scalar curvature singularity at $\tilde{y}=0$. The bulk spacetime is cut into two identical constant curvature parts by a hypersurface. The sharp point of this warp factor leads to a $\delta$-function in the second-order equations of motion for the general relativity case, which resembles the RS II brane world model. This case can be interpreted as the appearance of an exotic wall.

We will present the three solutions for the zero scalar curvature, dS, and AdS bulk spacetimes, respectively. As in the case of nonconstant curvature solutions, below we still retain the integration constant $c_1$ so that $c_1=0$ is the case satisfying Eqs.~\eqref{eq:1_MN} and \eqref{eq:2_pq}.

For the bulk spacetime with zero scalar curvature, there is a spinor field solution with the following $p(y)$ and $q(y)$
\begin{subequations}\label{eq:pq_curved_sol_f}
\begin{align}
  p(y)&=c_3 \exp\left(-\frac{c_1}{2 c_2}y+\frac{1}{c_2}y^{\frac{3}{5}}\right) y^{-\frac{4}{5}}\,, \\
  q(y)&=\frac{c_2}{c_3}\exp\left(\frac{c_1}{2 c_2}y-\frac{1}{c_2}y^{\frac{3}{5}}\right) y^{-\frac{4}{5}}\,,
\end{align}
\end{subequations}
and
\begin{align}\label{eq:Vphi_curved_f}
  V(\bar{\psi}\psi)=\frac{c_1}{2 c_2}\bar{\psi}\psi-
  \frac{12}{25}\left(\frac{\bar{\psi}\psi}{2 c_2}\right)^{\frac{5}{4}}\,.
\end{align}
An obvious property for this solution is that there is an additional nonlinear term in the potential. As shown in Fig.~\ref{subfig:rhoy_m_curved}, one can find that the energy density of the spinor field for this solution is infinite at the position $\tilde{y}=0$. Therefore, a scalar curvature singularity occurs at the position of $\tilde{y}=0$. The energy density approaches to zero as $\tilde{y}\rightarrow\pm\infty$. We could believe that an exotic spinor wall exists at $\tilde{y}=0$.

According to the definition~\eqref{eq:walltension}, the tension of the exotic wall in the domain $(-\frac{\sigma}{2},\frac{\sigma}{2})$ is
\begin{align}
  \varepsilon=\frac{5^{3/5} }{2^{1/5}}k (k \sigma )^{3/5}\,.
\end{align}
The tension of the exotic wall is not globally regular, but is locally regular.

For the dS bulk spacetime, a spinor field solution is given by
\begin{subequations}\label{eq:pq_curved_sol_dS}
\begin{align}
  p(y)&=c_3 \exp\left[-\frac{c_1}{2 c_2}y-H(y)\right] \cos^{-\frac{4}{5}}\left(\frac{5 k y}{2}\right)\,, \\
  q(y)&=\frac{c_2}{c_3}\exp\left[\frac{c_1}{2 c_2}y+H(y)\right] \cos^{-\frac{4}{5}}\left(\frac{5 k y}{2}\right)\,,
\end{align}
\end{subequations}
and
\begin{align}\label{eq:Vphi_curved_dS}
  V(\bar{\psi}\psi)=\frac{c_1}{2 c_2}\bar{\psi}\psi-6 k^2 \left[\left(\frac{\bar{\psi}\psi}{2 c_2}\right)^{\frac{5}{4}}-1\right]\,,
\end{align}
with the redefined function
\begin{align}\label{eq:function_G}
  H(y)=\frac{5 k}{2c_2} \cos ^{\frac{3}{5}}\left(\frac{5 k y}{2}\right) \mathrm{sgn}\left(\frac{5 k y}{2}\right)\, _2F_1\left(\frac{3}{10},\frac{1}{2};\frac{13}{10};\cos ^2\left(\frac{5 k y}{2}\right)\right)\,,
\end{align}
where $_2F_1$ is the hypergeometric function and is real. Therefore, $H(y)$ is a real function. As shown in Fig.~\ref{subfig:rhoy_dS_curved}, the energy density for this solution is valid for $\tilde{y}$ ranging from $-\frac{\pi}{5}$ to $\frac{\pi}{5}$, and mainly distributes in the domain $\left(-\frac{\pi}{5},\frac{\pi}{5}\right)$ with two divergence points of the energy density at the boundaries. This configuration can be viewed as a local trap.

The tension of the trap in the domain $(-\frac{\sigma}{2},\frac{\sigma}{2})$ is
\begin{align}
  \varepsilon= \frac{2 k}{5}\sin ^3\left(\frac{5 k \sigma }{4}\right) \, _2F_1\left(\frac{7}{10},\frac{3}{2};\frac{5}{2};\sin ^2\left(\frac{5 k \sigma }{4}\right)\right)\,.
\end{align}
The tension of the trap is not globally regular, but is locally regular.

For the AdS bulk spacetime, an analytic solution is obtained as
\begin{subequations}\label{eq:pq_curved_sol_AdS}
\begin{align}
  p(y)&=c_3 \exp\left[-\frac{c_1}{2 c_2}y-I(y)\right] \cosh ^{-\frac{4}{5}}\left(\frac{5 k y}{2}\right)\,, \\
  q(y)&=\frac{c_2}{c_3}\exp\left[\frac{c_1}{2 c_2}y+I(y)\right] \cosh ^{-\frac{4}{5}}\left(\frac{5 k y}{2}\right)\,,
\end{align}
\end{subequations}
and
\begin{align}\label{eq:Vphi_curved_AdS}
  V(\bar{\psi}\psi)=\frac{c_1}{2 c_2}\bar{\psi}\psi+6 k^2 \left[\left(\frac{\bar{\psi}\psi}{2 c_2}\right)^{\frac{5}{4}}-1\right]\,,
\end{align}
with the redefined function
\begin{align}\label{eq:function_H}
  I(y)=\mathrm{i}\frac{5 k}{2 c_2} \cosh^{\frac{3}{5}}\left(\frac{5 k y}{2}\right)\mathrm{sgn}\left(\frac{5 k y}{2}\right) \,_2F_1\left(\frac{3}{10},\frac{1}{2};\frac{13}{10};\cosh ^2\left(\frac{5 k y}{2}\right)\right)\,,
\end{align}
where the hypergeometric function $_2F_1$ is complex. Therefore, the redefined function $I(y)$ is complex. The imaginary component of this complex function should be discarded because it is introduced in the process of solving the equations, where the variable $y$ has been extended to a complex domain. As shown in Fig.~\ref{subfig:rhoy_AdS_curved}, the energy density increases and approaches to a constant as $\tilde{y}\rightarrow\pm\infty$. There is a trap of the energy density at $\tilde{y}=0$.

According to~\eqref{eq:walltension}, the tension of the trap in the domain $(-\frac{\sigma}{2},\frac{\sigma}{2})$ is
\begin{align}
  \varepsilon=&\frac{\left(3 k \text{csch}\left(\frac{5 k \sigma }{4}\right)\right) }{8\ 2^{3/10} \cosh ^{\frac{3}{5}}\left(\frac{5 k \sigma }{4}\right) \left(\cosh \left(\frac{5 k \sigma }{2}\right)+1\right)^{7/10}} \times\left[2 \cosh ^{\frac{13}{5}}\left(\frac{5 k \sigma }{4}\right) \left(\cosh \left(\frac{5 k \sigma }{2}\right)+4\right)\right.\nonumber\\&\left.-5 \left(\cosh \left(\frac{5 k \sigma }{2}\right)+1\right) \, _2F_1\left(-\frac{1}{2},\frac{7}{10};\frac{1}{2};-\sinh ^2\left(\frac{5 k \sigma }{4}\right)\right)\right]\,.
\end{align}
The tension of the trap is locally regular.

So far we have investigated spinor wall solutions with constant curvature. These solutions satisfy the first integral~\eqref{eq:first_integral_curved}. However, from the perspective of the energy density, the solution with zero scalar curvature leads to an exotic wall with an infinite large energy density at the position of the hypersurface. The energy density for the solution with zero scalar curvature has a singularity since the corresponding bulk spacetime has a scalar curvature singularity at $\tilde{y}=0$. The energy density for the dS solution is not bounded from above and this solution is not globally regular. The energy density for the AdS solution converges to a constant as $\tilde{y}$ tends to the boundary and this solution is also not globally regular. Nonetheless, from the local perspective, the energy densities of the dS and AdS solutions have local traps. Moreover, the stability of the above walls and the localization of gravitation need to be considered.

\section{Stability of spinor walls}
\label{sec:stability}

Having found spinor wall solutions, we are obliged to examine perturbations about this background with the purpose of studying their stability. In general, according to the four-dimensional Poincar\'{e} transformations, the linear perturbations of the metric~\eqref{eq:line_element_curved} and the energy–momentum tensor can be mathematically decomposed into transverse-traceless (TT) tensor mode, transverse vector modes, and scalar modes. In our case, the perturbation of the metric will couple to the perturbation of the bulk spinor field. The spinor field here behaves like a scalar field in spacetime. Thus, the perturbation of the bulk spinor field is similar to the perturbation of a scalar field. As a consequence, the stability of the spinor walls can be analyzed with the help of methods in~\cite{Giovannini:2001gg,Giovannini:2002ga,Giovannini:2002gb,Giovannini:2003g}.

For the sake of convenience, we introduce the transformation $\mathrm{d}z=\mathrm{d}y/a(y)$ for the metric~\eqref{eq:line_element_curved}.
The linear perturbation to the metric~\eqref{eq:line_element_curved} in the conformal coordinate $z$ can be parameterized as
\begin{align}\label{eq:perturbation}
\delta^{(1)} g_{MN}=a^2(z)
\begin{pmatrix}
2h_{\mu\nu}+\left(\partial_\mu \xi_\nu+ \partial_\nu \xi_\mu\right) +2\eta_{\mu\nu}\theta+
2\partial_\mu \partial_\nu \omega &  \zeta_\mu +\partial_\mu \phi  \\
\zeta_\nu +\partial_\nu \phi &  2 \chi
\end{pmatrix}\,,
\end{align}
where $\delta^{(1)}$ signifies linear perturbation. Its inverse will be
\begin{align}
\delta^{(1)} g^{MN}=\frac{1}{a^2(z)}
\begin{pmatrix}
-2h^{\mu\nu}-\left(\partial^\mu \xi^\nu+ \partial^\nu \xi^\mu\right) -2\eta^{\mu\nu}\theta-
2\partial^\mu \partial^\nu \omega &  \zeta^\mu +\partial^\mu \phi  \\
\zeta^\nu +\partial^\nu \phi &  -2 \chi
\end{pmatrix}\,.
\end{align}
The TT tensor mode $h_{\mu\nu}$ satisfies
\begin{align}\label{eq:TT}
  \partial^\mu h_{\mu\nu}=0\,,\quad h=\eta^{\mu\nu}h_{\mu\nu}=0\,.
\end{align}
The vector modes $\xi_{\mu}$ and $\zeta_\mu$ are both transverse
\begin{align}\label{eq:T}
  \partial^\mu \xi_{\mu}=0\,,\quad \partial^\mu \zeta_\mu=0\,.
\end{align}
The four components $\theta$, $\omega$, $\phi$, and $\chi$ in Eq.~\eqref{eq:perturbation} are scalar modes of the metric perturbation.

The vielbeins can be separated into their background and perturbation parts:
\begin{align}
\tilde{e}^{\bar{N}}_{\;M}=e^{\bar{N}}_{\;M}+\delta^{(1)} e^{\bar{N}}_{\;M}\,.
\end{align}
The unperturbed vielbeins and its inverse are respectively
\begin{align}
e^{\bar{N}}_{\;M}=a(z)
\begin{pmatrix}
\delta_{\;\mu}^{\bar{\nu}}  & 0  \\ 0 &  1
\end{pmatrix}\,,
\quad\
e^{\;M}_{\bar{N}}=\frac{1}{a(z)}
\begin{pmatrix}
\delta^{\;\mu}_{\bar{\nu}}  & 0  \\ 0 &  1
\end{pmatrix}\,.
\end{align}
According to Eq.~\eqref{eq:perturbation}, the corresponding perturbation part of the vielbeins~\footnote{The scalar-vector-tensor decomposition of the perturbations is similar to the cosmological version~\cite{Golovnev:2018wbh}.} can be parameterized as
\begin{align}
\delta^{(1)} e^{\bar{N}}_{\;M}=a(z)
\begin{pmatrix}
\delta^{\bar{\nu}}_{\;\gamma} \left[h^{\gamma}_{\mu}+\partial_{\mu} \xi^{\gamma} +\delta^{\gamma}_{\mu}\theta+ \partial^{\gamma} \partial_{\mu} \omega + \eta^{\gamma\alpha} \epsilon_{\alpha\mu\iota} \left(\zeta_3^{\iota} +\partial^{\iota} \phi_3\right)\right] & \delta^{\bar{\nu}}_{\;\gamma} \left(\zeta_1^{\gamma} +\partial^{\gamma} \phi_1\right)  \\  \delta_{\;\mu}^{\bar{\gamma}} \left(\zeta_{2\bar{\gamma}} +\partial_{\bar{\gamma}} \phi_2\right) &  \chi
\end{pmatrix}\,,
\end{align}
and the inverse is
\begin{align}
\delta^{(1)} e^{\;M}_{\bar{N}}=\frac{-1}{a(z)}
\begin{pmatrix}
\delta^{\;\mu}_{\bar{\gamma}} \left[h^{\bar{\gamma}}_{\bar{\nu}}+\partial_{\bar{\nu}} \xi^{\bar{\gamma}} +\delta^{\bar{\gamma}}_{\bar{\nu}}\theta+ \partial^{\bar{\gamma}} \partial_{\bar{\nu}} \omega+ \eta^{\bar{\gamma}\bar{\alpha}} \epsilon_{\bar{\alpha}\bar{\nu}\bar{\iota}} \left(\zeta_3^{\bar{\iota}} +\partial^{\bar{\iota}} \phi_3\right) \right] & \delta^{\;\mu}_{\bar{\gamma}} \left(\zeta_1^{\bar{\gamma}} +\partial^{\bar{\gamma}} \phi_1\right)  \\ \delta_{\bar{\nu}}^{\;\gamma} \left(\zeta_{2\gamma} +\partial_{\gamma} \phi_2\right) &  \chi
\end{pmatrix}\,,
\end{align}
where $\epsilon_{\alpha\beta\gamma}$ is the permutation tensor, $\phi_3$ is a pseudoscalar, $\zeta_3^{\iota}$ is a pseudovector, and
\begin{align}
  \zeta_{1\mu}+\zeta_{2\mu}=\zeta_{\mu} \,, \quad \phi_1+\phi_2=\phi \,.
\end{align}

The spinor field can be separated into its background and perturbation parts:
\begin{align}
  \tilde{\psi}= \psi+ \delta^{(1)}\psi\,.
\end{align}
The evolution equations of the linear perturbations are
\begin{align}\label{eq:EOM of perturbations}
  \delta^{(1)} R_{MN}-\frac{1}{2}\delta^{(1)} R \bar{g}_{MN}-\frac{1}{2} \bar{R} \delta^{(1)} g_{MN}&=\delta^{(1)} T_{MN}\,, \\
  \delta^{(1)}\gamma^{M} \nabla_{M} \psi+ \gamma^{M} \delta^{(1)} \nabla_{M} \psi+\frac{\partial^2 V}{\partial \bar{\psi} \partial \psi} \delta^{(1)} \psi &=0 \,, \label{eq:EOM1 of perturbations_spinor} \\
  \delta^{(1)}\nabla_M\bar{\psi}\gamma^{M}+\nabla_M\bar{\psi}\delta^{(1)}\gamma^{M}
  -\frac{\partial^2 V}{\partial \psi\partial \bar{\psi}}\delta^{(1)} \bar{\psi} &=0\,. \label{eq:EOM2 of perturbations_spinor}
\end{align}
and the perturbation of the energy-momentum tensor is given by
\begin{align}
\delta^{(1)} T_{M N}=& \frac{1}{2}\left[\delta^{(1)} \bar{\psi} \gamma_{(M} \nabla_{N)} \psi+\bar{\psi} \delta^{(1)} \gamma_{(M} \nabla_{N)} \psi+\bar{\psi} \gamma_{(M} \delta^{(1)} \nabla_{N)} \psi-\delta^{(1)} \nabla_{(M} \bar{\psi} \gamma_{N)} \psi \right. \nonumber \\
&\left. -\nabla_{(M} \bar{\psi} \delta^{(1)} \gamma_{N)} \psi-\nabla_{(M} \bar{\psi} \gamma_{N)} \delta^{(1)}\psi\right]+ \delta^{(1)} g_{M N}\left[\frac{1}{2}\left(\bar{\psi} \frac{\partial V}{\partial \bar{\psi}}+\frac{\partial V}{\partial \psi} \psi\right)-V\right] \nonumber \\
& +g_{M N} \frac{1}{2}\left[\left(\frac{\partial^2 V}{\partial \bar{\psi} \partial \psi} \bar{\psi} \delta^{(1)} \psi +\frac{\partial^2 V}{\partial \psi \partial \bar{\psi}} \delta^{(1)}\bar{\psi} \psi\right) - \left(\delta^{(1)}\bar{\psi}\frac{\partial V}{\partial \bar{\psi}}+\frac{\partial V}{\partial \psi} \delta^{(1)}\psi\right)\right]\,.
\end{align}
Here, the perturbations of the gamma matrices and the covariant derivative of the spinor field $\psi$ are
\begin{align}
\delta^{(1)}\gamma^{M} =&\delta^{(1)} e_{\bar{N}}^{M} \gamma^{\bar{N}} \,, \\
\delta^{(1)} \nabla_{M} \psi =&\partial_{M} \delta^{(1)} \psi+\frac{\mathrm{i}}{2} \delta^{(1)} \omega_{M}^{\bar{A}\bar{B}} S_{\bar{A}\bar{B}} \psi+\frac{\mathrm{i}}{2} \omega_{M}^{\bar{A}\bar{B}} S_{\bar{A}\bar{B}} \delta^{(1)} \psi \,,
\end{align}
respectively. Using Eq.~\eqref{eq:spin_connection} the spin connection in the linear perturbation can be computed as:
\begin{align}
  \delta^{(1)} \omega_{M}^{\bar{A}\bar{B}}=&
  \frac{1}{2}\delta^{(1)}e^{N\bar{A}}\left(\partial_{M}e^{\bar{B}}_{N}-\partial_{N}e^{\bar{B}}_{M}\right)+
   \frac{1}{2}e^{N\bar{A}}\delta^{(1)}\left(\partial_{M}e^{\bar{B}}_{N}-\partial_{N}e^{\bar{B}}_{M}\right)\nonumber \\
 & -\frac{1}{2}\delta^{(1)}e^{N\bar{B}}\left(\partial_{M}e^{\bar{A}}_{N}-\partial_{N}e^{\bar{A}}_{M}\right)-
   \frac{1}{2}e^{N\bar{B}}\delta^{(1)}\left(\partial_{M}e^{\bar{A}}_{N}-\partial_{N}e^{\bar{A}}_{M}\right)
 \nonumber \\
 &-\frac{1}{2}\delta^{(1)}e^{P\bar{A}}e^{Q\bar{B}}\left(\partial_{P}e_{Q\bar{C}}-\partial_{Q}e_{P\bar{C}}\right)e^{\bar{C}}_{M}
 -\frac{1}{2}e^{P\bar{A}}\delta^{(1)}e^{Q\bar{B}}\left(\partial_{P}e_{Q\bar{C}}-\partial_{Q}e_{P\bar{C}}\right)e^{\bar{C}}_{M}\nonumber \\
 &-\frac{1}{2}e^{P\bar{A}}e^{Q\bar{B}}\delta^{(1)}\left(\partial_{P}e_{Q\bar{C}}-\partial_{Q}e_{P\bar{C}}\right)e^{\bar{C}}_{M}
 -\frac{1}{2}e^{P\bar{A}}e^{Q\bar{B}}\left(\partial_{P}e_{Q\bar{C}}-\partial_{Q}e_{P\bar{C}}\right)\delta^{(1)}e^{\bar{C}}_{M}\,.
\end{align}
The nonvanishing components of the perturbed spin connection are
\begin{align}
  \delta^{(1)} \omega_{\mu}^{\bar{\alpha}\bar{\beta}}=& \frac{1}{2} \left(a^2-1\right)\left(\partial^{\bar{\alpha}}h^{\bar{\beta}}_{\mu}-\partial^{\bar{\beta}}h^{\bar{\alpha}}_{\mu}\right) \,,\\
  \delta^{(1)} \omega_{\mu}^{\bar{5}\bar{\beta}}=& \delta^{(1)} \omega_{\mu}^{\bar{\beta}\bar{5}}= \frac{1}{2}\partial_{5}\left(a h^{\bar{\beta}}_{\mu}\right) \,.
\end{align}
For Eqs.~\eqref{eq:EOM1 of perturbations_spinor} and \eqref{eq:EOM2 of perturbations_spinor}, the equation of motion of the tensor perturbation leads to the following constrains
\begin{align}\label{eq:constrain}
  \delta^{(1)} e_{\bar{\nu}}^{\mu} \gamma^{\bar{\nu}} \nabla_{\mu} \psi+ e_{\bar{\nu}}^{\mu} \gamma^{\bar{\nu}} \delta^{(1)} \omega_{\mu} \psi=0 \,, \\
  \nabla_{\mu}\bar{\psi}\delta^{(1)}e_{\bar{\nu}}^{\mu}\gamma^{\bar{\nu}}- \bar{\psi}\delta^{(1)}\omega_{\mu}e_{\bar{\nu}}^{\mu}\gamma^{\bar{\nu}}=0\,.
\end{align}

In fact, the perturbation $\delta^{(1)}\psi$ of the spinor field $\psi$ will contribute to the scalar modes of the perturbation. A complete investigation of the stability of this system includes tensor mode, vector modes, and scalar modes. For the sake of simplicity, we only consider the tensor perturbation here. As a consequence, the perturbation of the energy-momentum tensor is vanishing for the tensor perturbation. We eliminate the gauge degrees of freedom not by choosing a gauge but by working only with gauge-invariant quantities. The tensor mode $h_{\mu\nu}$ is gauge-invariant~\cite{Giovannini:2001gg}. We obtain the following tensor perturbation equation
\begin{align}\label{eq:EOM_tensor}
  \left[\square^{(4)}+\partial_z^{2}+\left(3\frac{\partial_za}{a}\right)\partial_z\right]h_{\mu\nu}(x^{\mu},z)&=0\,,
\end{align}
where $\square^{(4)}=\eta^{\mu\nu}\partial_\mu\partial_\nu$ is the four-dimensional d'Alembert operator.

After performing the separation of variables, the equation of motion of the tensor mode is separated into the Klein-Gordon equation for the four-dimensional part and a second-order differential equation for the extra-dimensional part. For the extra-dimensional part of the tensor mode, we can transform it into a canonically normal mode and obtain a Schr\"{o}dinger-like equation for the extra-dimensional part~\cite{Giovannini:2002gb}
\begin{align}\label{eq:Schrodinger-like}
  \left[-\partial_z^2+W(z)\right]\Psi(z)=m^2\Psi(z)\,.
\end{align}
Here, the effective potential $W(z)$ has the following form
\begin{align}\label{eq:eff_potential_z}
W(z)=\Omega^2(z)+\partial_z\Omega(z)
\end{align}
with
\begin{equation}\label{eq:zOmega}
\Omega(z)=\frac{1}{2}\partial_z \ln a^3(z)\,.
\end{equation}
Consequently, the Schr\"{o}dinger-like equation~\eqref{eq:Schrodinger-like} can be factorized as a supersymmetric quantum mechanics form
\begin{align}\label{eq:SUSY-form_partner1}
  \mathcal{Q}(z)\,\mathcal{Q}^{\dagger}(z)\psi(z)=m^2\psi(z)
\end{align}
with
\begin{equation}
\mathcal{Q}(z)=\partial_z+\Omega(z)\,,\quad \mathcal{Q}^{\dagger}(z)=-\partial_z+\Omega(z)\,,
\end{equation}
where the ``Hamiltonian'' $\mathcal{Q}(z)\,\mathcal{Q}^{\dagger}(z)$ is a positive definite Hermitian operator and there are no normalizable negative energy graviton modes.

The supersymmetric form of the Schr\"{o}dinger-like equation demonstrates that any tachyonic modes with $m^2<0$ are absent for the tensor mode. Therefore, these spinor walls can exist stably under the tensor perturbation in such spacetime with the metric~\eqref{eq:line_element_curved}. It is worth stressing that this result is valid only for a regular bulk spacetime without singularities in curvature invariance. From this perspective, the stability of the spinor wall solution in the spacetime with zero scalar curvature is still unknown. The spinor wall solution in the spacetime with zero scalar curvature represents an exotic wall which has a singularity at the location of the wall.

\section{Localization of gravitation}
\label{sec:localization}

In what follows, we will discuss localization of the zero mode of the tensor perturbation for different warp factors investigated in section~\ref{sec:model}. In the process of obtaining a four-dimensional effective gravitational theory from high dimensions, a four-dimensional effective coupling constant could be obtained. To quadratic order in the tensor perturbation, the gravitational action would be
\begin{align}
  S_{\mathrm{g}}=\frac{M_{(5)}^3}{2}\int\mathrm{d}^5x\left[\delta^{(2)}\sqrt{-g^{(5)}}R
  +\sqrt{-g^{(5)}}\delta^{(2)}R+\delta\sqrt{-g^{(5)}}\delta R\right]\,,
\end{align}
where $\delta^{(2)}$ denotes second-order quantities in the tensor perturbation. Taking the TT condition~\eqref{eq:TT} and the separation of variables in the coordinates $\left(x^{\mu}, z\right)$ into account, one obtain a four-dimensional effective theory with the Einstein-Hilbert action and additional terms:
\begin{align}\label{eq:eff_action}
  S_{\textrm{eff}}\supset\frac{M_{(5)}^3}{2}\int\mathrm{d}^4x\left[-\frac{1}{4}\eta^{\alpha\beta}\partial_\alpha \epsilon^{\rho\lambda}\left(x^{\mu}\right)\partial_\beta \epsilon_{\rho\lambda}\left(x^{\mu}\right)\right] \int\mathrm{d}z\,\Psi^2(z)\,.
\end{align}
The four-dimensional part of such canonical action is anticipated to describe gravitation observed experimentally. The four-dimensional effective Planck mass $M_{\textrm{Pl}}$ is expressed as follows
\begin{align}\label{eq:Planck}
  M_{\textrm{Pl}}^2=M_{(5)}^{3} \int_{\Sigma}\mathrm{d}z\,\Psi^2(z)\,.
\end{align}
Here, $\Sigma$ is the integration interval of the fifth dimension $z$. For an infinite extra dimension, the integration interval will be $\left(-\infty,\infty\right)$. An arbitrary KK mode is normalizable if and only if the following normalization condition is satisfied
\begin{align}\label{eq:normalization_condition}
  \int_{\Sigma}\Psi^2(z) \,\mathrm{d}z<\infty\,.
\end{align}
In order to obtain the four-dimensional Newtonian potential, the tensor zero mode is required to be localized.

From Eq.~\eqref{eq:Schrodinger-like}, we obtain a zero eigenvalue solution~\cite{Giovannini:1997g,Cui:2020clwlz}, namely the zero mode of the graviton
\begin{align}\label{eq:zero_mode}
  \Psi_0(z)=\left[a^3(z)\right]^{1/2}\left[C_1+ C_2 \int
  \frac{1}{a^3(z)}\mathrm{d}z\right]\,.
\end{align}
Here, $C_1$ and $C_2$ are integration constants. There exists the possibility of the normalization of the zero mode with this general solution.

Furthermore, the graviton zero mode propagating in the bulk is subjected to boundary conditions~\cite{Gherghetta:2010}. For our case, the vanishing variation of the action of the zero mode at the boundary $\partial\Sigma$ of the fifth dimension $z$ leads to the boundary condition
\begin{align}\label{eq:boundary_condition}
  \delta h^{\mu\nu}\partial_z h_{\mu\nu}\big|_{\partial\Sigma}=0\,.
\end{align}
Notice that $h_{\mu\nu}$ is assumed to vanish at the four-dimensional boundary $x^{\rho}\rightarrow \pm \infty$, and corresponding boundary terms are automatically zero. Accordingly, the boundary condition~\eqref{eq:boundary_condition} can again be satisfied by imposing either
\begin{align}
  \text{Dirichlet condition:}& \quad \left[a^3(z)\right]^{-1/2}\Psi_0\left(z\right)\bigg|_{\partial\Sigma}=0\,,\label{eq:Dirichlet_condition}\\
  \text{or Neumann condition:}& \quad \partial_z\left\{\left[a^3(z)\right]^{-1/2} \Psi_0\left(z\right)\right\}\bigg|_{\partial\Sigma}=0\,.\label{eq:Neumann_condition}
\end{align}
The zero eigenvalue solution of the Schr\"{o}dinger-like equation~\eqref{eq:Schrodinger-like} is required to satisfy these boundary conditions.

The Dirichlet condition allows the zero mode with the second particular solution
\begin{align}\label{eq:second_particular_sol}
  \Psi_0(z)=C_2\left[a^3(z)\right]^{1/2} \int \frac{1}{a^3(z)}\mathrm{d}z\,.
\end{align}
The Neumann condition leads to the zero mode with the general solution~\eqref{eq:zero_mode}. Especially, the first particular solution
\begin{align}\label{eq:first_particular_sol}
  \Psi_0(z)=C_1\left[a^3(z)\right]^{1/2}
\end{align}
identically meets the Neumann condition, which is independent of the concrete form of the warp factor. From this result, the solutions of the warp factor that we have investigated above coincide with the requirement of the Neumann condition. Essentially, one does not rule out the possibility that the second particular solution~\eqref{eq:second_particular_sol} satisfies the boundary condition~\eqref{eq:boundary_condition} for some warp factors. In the following, we discuss the localization of the graviton zero mode only for the first particular solution~\eqref{eq:first_particular_sol}.

For an AdS bulk or an asymptotically AdS one, the warp factors may be exponentially divergent at the fifth-dimensional boundary $z\rightarrow \pm \infty$ and the first particular solution~\eqref{eq:first_particular_sol} presumably is not normalizable. Thus, only the (asymptotically) properties of the bulk geometry are not sufficient. The normalization of the first particular solution~\eqref{eq:first_particular_sol} will be discussed in the following. Inserting~\eqref{eq:first_particular_sol} into the normalization condition~\eqref{eq:normalization_condition}, one obtains
\begin{align}\label{eq:normalization}
  \int_{\Sigma}\Psi_0^2(z) \,\mathrm{d}z=\int_{\Sigma}a^3(z)\,\mathrm{d}z=\int_{\Sigma'}a^2(y)\,\mathrm{d}y<\infty\,.
\end{align}
Here, $\Sigma'$ is the integration interval in the $y$ coordinate.

For the warp factor~\eqref{eq:warp_factor1}, we have
\begin{align}
  \int_{-\infty}^{\infty}\text{sech}^{2n}(y)\,\mathrm{d}y=\frac{4^n}{k n}\,_2F_1(n,2 n;n+1;-1)\,.
\end{align}
As a result, if $n>0$ the integral~\eqref{eq:normalization} is convergent and the zero mode of the graviton~\eqref{eq:first_particular_sol} is normalized. The warp factor in the Janus solution~\eqref{eq:Janus} or \eqref{eq:warp_factor_Janus} leads to a divergent result of the integral~\eqref{eq:normalization} and the graviton zero mode is not localized. We also conclude that for the warp factor $\cosh^{\frac{2}{5}}\left(\frac{5}{2}ky\right)$ corresponding to the AdS constant curvature spacetime, the graviton zero mode is not localized. For the warp factor~\eqref{eq:warp_factor2}, the result will be
\begin{align}
  \int_{-\frac{\pi}{2k}}^{\frac{\pi}{2k}}\cos^{2n}(y)\,\mathrm{d}y=\frac{\sqrt{\pi}}{k}\frac{\Gamma \left(n+\frac{1}{2}\right)}{\Gamma (n+1)}\,.
\end{align}
Consequently, if $n>-\frac{1}{2}$ the integral~\eqref{eq:normalization} is convergent and there is a normalized graviton zero mode. Therefore, the warp factor for the spinor wall solution with the dS bulk spacetime achieves a normalizable graviton zero mode. We conclude that for the warp factor $\cos^{\frac{2}{5}}\left(\frac{5}{2}ky\right)$ corresponding to the dS constant curvature spacetime, the graviton zero mode is localized. For the warp factor~\eqref{eq:warp_factor3}, the result is
\begin{align}
  \int_{-\infty}^{\infty}\exp\left[-2(k y)^2\right]\,\mathrm{d}y=\frac{1}{| k|}\sqrt{\frac{\pi }{2}}\,.
\end{align}
In this case, the zero mode of the graviton is localized. From these results, spinor walls with warp factors~\eqref{eq:warp_factor1}, \eqref{eq:warp_factor2}, and \eqref{eq:warp_factor3} can localize the graviton zero mode under specific conditions.

To summarize, under our assumptions about the background, it is shown that the gauge-invariant perturbation corresponding to the tensor mode of the geometry is localized on the walls. The localized zero mode of the tensor perturbation results in the four-dimensional general relativity and hence the four-dimensional Newtonian potential. The massive KK modes of the tensor perturbation will make a correction $\Delta U(r)\sim1/r^2$ concerning the Newtonian potential~\cite{Randall:1999rsa,Csaki:2000cehs}.

\section{Discussion and conclusions}
\label{sec:conclusions}

In the above solutions, we found the solution corresponding to the warp factor~\eqref{eq:warp_factor1} with $n>0$ satisfies the conditions that the energy density is regular; the solution is stable under the tensor perturbation; and the graviton zero mode can be localized on the wall. For the warp factor~\eqref{eq:warp_factor1}, there exists another solution, for which the energy density is regular; the solution is stable under the tensor perturbation; while the graviton zero mode cannot be localized on the wall. The solution corresponding to the warp factor~\eqref{eq:warp_factor_Janus} also belongs to this case. Moreover, we found an interesting result that a global trap can stably exist, but its energy density is not regular and the graviton zero mode cannot be localized on the trap. For the solutions with the warp factors~\eqref{eq:warp_factor2} and \eqref{eq:warp_factor3}, we believe that local walls or traps can exist stably, whether or not the graviton zero mode is localized. The constant curvature solutions in the dS and AdS bulk spacetimes are special cases of the solutions with the warp factors~\eqref{eq:warp_factor3} and \eqref{eq:warp_factor1} respectively. The constant curvature solution in the bulk spacetime with zero scalar curvature corresponds to an exotic wall.

We comment on the divergences in the above cases. For the constant curvature solution in the bulk spacetime with zero scalar curvature, the wall which has a curvature singularity at the origin can be considered as a thin wall. For naked singularities at the boundaries of the bulk, the walls/traps can be physically meaningful. One reason is that in renormalization group flows to nonconformal theories, the AdS horizon is replaced by a naked singularity. This singularity is physical in the sense that the singular behavior corresponds to strong coupling effects such as confinement or screening in the boundary theory. The other reason comes from analyzing the spectrum of gravity from a four-dimensional point of view. The KK states couple to matter on the thin domain wall and cause small violations of the four-dimensional conservation of energy and momentum. Spaces with naked singularities are physically acceptable only if one imposes boundary conditions that guarantee the four-dimensional conservation of energy and momentum.

In summary, we have investigated analytic solutions of a spinor field in five-dimensional spacetime. We mainly focused on analytic solutions of the spinor field for flat and warped spacetimes. First, we considered a toy model in which a real spinor field provides an illuminating solution. We found a set of solutions in the absence of gravitation and three sets of solutions of the spinor field in the presence of gravitation. These solutions with gravitation are discussed for three kinds of warp factors which correspond to different warped geometries. Further, we discussed domain wall solutions of the spinor field in the five-dimensional zero scalar curvature, dS, and AdS bulk spacetimes. In order to identify stability of these walls, we studied perturbations of these wall systems. The proof demonstrate that these spinor walls are stable under the tensor perturbation. Moreover, it was shown that the zero mode of the graviton can be localized on the walls in the five-dimensional bulk spacetime with nonconstant curvature. It implies that the Newtonian potential is recovered.

In many works in which analytical solutions of domain walls or thick branes were found, scalar fields were widely considered because not only it is easy to be solved but also matter fields can be localized by employing a natural mechanism. In this work, we attempted to explore the possibility of domain walls generated by a real spinor field. We found that, if such field exists, appropriate choices of the spinor potential would generate a rich variety of behaviours, quite different from their widely studied scalar field counterparts. The energy density of the spinor field concentrates in the vicinity of walls except for traps. Moreover, nonlinear spinor fields may shed light on torsion of the bulk spacetime, and present a torsionless four-dimensional hypersurface which may stands for a world with the illusion of torsion free~\cite{Mukhopadhyaya:2002mss}.

Whether these spinor wall solutions can be seen as branes or not is yet important to verify. Branes require that matter fields are confined on them or bounded in the vicinity of them. Therefore, the dynamics of matter fields in the presence of a wall configuration is also a topic worth studying. In this work, we studied solutions for a real spinor field. However, from the perspective of fundamental particles, fermions in the standard model are described by complex spinors with four components. It is interesting to investigate solutions for a complex spinor field in warped five-dimensional geometries. So far, we only considered a simple Lorentz invariant $\bar{\psi}\psi$ for a spinor field in our solutions. However, other Lorentz invariants in the potential term $V(\bar{\psi}, \psi)$ of the Lagrangian density, such as $\bar{\psi}\gamma^5\psi$, current-current interaction density $\left(\bar{\psi}\gamma^\mu\psi\right)\left(\bar{\psi}\gamma_\mu\psi\right)$, and $\left(\bar{\psi} \gamma^{5} \gamma^{\mu} \psi\right)\left(\bar{\psi} \gamma^{5} \gamma_{\mu} \psi\right)$, are also rational. We will leave this in future works. In this work, we only considered a wall with the four-dimensional Poincar\'{e} symmetry. Naturally, it is worth exploring other walls which possess other symmetries, such as the four-dimensional dS or AdS symmetry.

\section*{Acknowledgements}

The authors would like to thank Tao-Tao Sui for helpful discussions and Simon F. Ross for helpful comments. This work was supported in part by the National Natural Science Foundation of China (Grants No. 11875151, No. 12047501, and No. 12247135) and Lanzhou City's scientific research funding subsidy to Lanzhou University. Zheng-Quan Cui was funded by the China Scholarship Council.

\bibliographystyle{JHEP}
\bibliography{ref}

\end{document}